\documentclass{JHEP3}

\usepackage{epsfig}

\renewcommand\t{\theta}
\renewcommand\l{\lambda}
\renewcommand\a{\alpha}
\renewcommand\b{\beta}
\newcommand\e{\ensuremath{\epsilon}}
\newcommand\R[1]{\ensuremath{\mathbf{#1}}}
\newcommand\Rb[1]{\ensuremath{\overline\mathbf{#1}}}
\newcommand\tb{\ensuremath{\mathop{\rm tan}\beta}}
\newcommand\F{{\ensuremath{\cal F}}}

\renewcommand\Re{\ensuremath{\mathop{\rm Re}}}

\newcommand\eV{\mbox{eV}}
\newcommand\MeV{\mbox{MeV}}
\newcommand\GeV{\mbox{GeV}}
\newcommand\TeV{\mbox{TeV}}

\newcommand\Md{\ensuremath{m_d}}
\newcommand\Ms{\ensuremath{m_s}}
\newcommand\be{\begin{equation}}
\newcommand\ee{\end{equation}}
\newcommand\bea{\begin{eqnarray}}
\newcommand\eea{\end{eqnarray}}
\newcommand\ba{\begin{array}}
\newcommand\ea{\end{array}}
\newcommand\bma{\begin{array}{ccc}}
\newcommand\ema{\end{array}}
\newcommand\matr[1]{\left(\bma#1\ema\right)}
\newcommand\0{\nonumber}

\newcommand{\SU}{\mathop{\rm SU}\nolimits}

\title{Supersymmetric SO(10) for fermion masses and mixings:
       rank-1 structures of flavour}

\author{Zurab Berezhiani and Fabrizio Nesti\\
  Dipartimento di Fisica, Universit\`a dell'Aquila\\
67010 Coppito, L'Aquila, and\vspace*{1ex}\\
  INFN, Laboratori Nazionali del Gran Sasso\\
67100 Assergi, L'Aquila, Italy\vspace*{ .5ex}\\
E-mail: \email{zurab.berezhiani@aquila.infn.it}, \email{fabrizio.nesti@aquila.infn.it}}

\abstract{We consider a supersymmetric SO(10) model with a SU(3) symmetry of
flavour in which fermion masses emerge via the see-saw mixing with
superheavy fermions in $16+\overline{16}$ representations. In this
model the dangerous $D=5$ operators of proton decay are naturally
suppressed and flavour-changing supersymmetric effects are under
control.  The mass matrices for all fermion types (up and down quarks,
charged leptons as well as neutrinos) appear in the form of
combinations of three rank-1 matrices, common to all types of
fermions, with different coefficients that are successive powers of
small parameters, related to each other by SO(10) symmetry properties.
Two versions of the model are considered, in which approximate grand
unification of masses takes place between quarks and leptons of the
first family (with very small $\tan\beta$) or for the ones of the
second family (predicting moderate $\tan\beta \simeq 7$--$8$).
The second version exhibits an interesting mechanism of unification of
the determinants of the Yukawa matrices of all types of fermions at
the GUT scale and it provides a perfect fit of the known data
for fermion masses, mixing and CP-violation.
It predicts a hierarchical pattern of neutrino masses with non-zero
$\theta_{e3}$, within 2-7 degrees. In addition, it predicts the
correct sign of the baryon asymmetry of the Universe via the
leptogenesys scenario.}


\begin{document}

\section{Introduction}

In the recent years the experimental information on fermion masses and
mixing has become increasingly accurate, with hierarchies and apparent
regularities representing a puzzle that continue to call for new
theoretical frameworks beyond the Standard Model. Among these, the
most promising possibility is related to the minimal supersymmetric
extension of the Standard Model (MSSM). In the MSSM, the fermion
masses and mixings emerge from the Yukawa couplings to the Higgs
fields $h_u$, $h_d$: 
\be
q \hat\l_u u^c\,h_u\,+\,q \hat\l_d d^c\,h_d+l\hat\l_e e^c\,h_d+l\hat\l_{\nu_D}\nu^c\,h_u\,,
\label{eq:SMyuk}
\ee
 where $q=(u,d)$, $l=(\nu,e)$ are weak isodoublets and $u^c$, $d^c$,
$e^c$, $\nu^c$ are isosinglets (the family indices are suppressed) and
$\hat\l$'s are matrices in family space.\footnote{Throughout the paper
we will represent quantities that are 3$\times$3 matrices of flavour
with a hat, i.e.~$\hat \l=\{\l_{ij}\}$, and denote flavour indices
with $i$, $j$.}  Also the neutrino Yukawa couplings are included that
induce their Dirac mass terms. However since $\nu^c$ are neutral, they
can have Majorana mass terms ${\cal M}\, \nu^c\hat \l_{\nu^c}\nu^c$ where
$\cal M$ is a large mass scale related to some physics beyond the
Standard Model.  As a result the effective D=5 operators are induced
by the seesaw mechanism~\cite{seesaw}:
\be
\frac{\hat\l_{\nu}}{\cal M}\, l l \, h_uh_u\,,
\qquad\qquad
\hat\l_{\nu}=\hat\l_{\nu_D}\hat\l_{\nu^c}^{-1}\hat\l_{\nu_D}^t\,.
\ee

Regarding charged fermions, we observe that their Yukawa couplings
(eigenvalues of the matrices $\hat\l_{u,d,e}$) show strong hierarchy
in family space:
\bea
\l_u\ :\ \l_c  \ :\ \l_t   &\ \sim\ &\e_u^2\ :\ \e_u\ :\ 1\qquad\e_u\sim1/500\0\\
\l_d\ :\ \l_s  \ :\ \l_b   &\ \sim\ &\e_d^2\ :\ \e_d\ :\ 1\qquad\e_d\sim1/50\0\\
\l_e\ :\ \l_\mu\ :\ \l_\tau&\ \sim\ &\e_e^2\ :\ \e_e\ :\ 1\qquad\e_e\sim1/50
\label{eq:up}
\eea
while the quarks mixing angles are small: $s_{12}\sim\sqrt{\e_d}$,
$s_{23}\sim\e_d$, $s_{13}\sim\e_d^2$.  In the ``up'' sector (U) the
scaling pattern~(\ref{eq:up}) is almost exact while in the ``down''
and ``charged leptons'' sectors (D, E) there are sensible deviations.
The deviations are present mainly for the first two generations in the
D, E sectors, and we will naturally connect this with the largeness of
the Cabibbo mixing angle, as compared to the other quark mixings. From
the above list we note that the ``down'' and ``charged leptons''
sectors have similar hierarchies, reaching the well known observation
that $\e_u\ll\e_e\simeq\e_d$.

On the other hand, in the neutrino sector, the neutrino mass
eigenstates show a milder hierarchy (they could also be strongly
degenerate) while the mixing angles are large.

The concept of grand unified models, together with the idea of family
unification provides a promising framework for solving the flavour
problems.  In this respect, probably the SO(10) is the most
interesting candidate, since it unifies all the fermions of one family
into a single multiplet $\R{16}$ and so it can naturally link the
Yukawa constants of different fermion sectors, with typical Clebsch
factors of SO(10)~\cite{so10}.  For family unification the maximal
symmetry group is SU(3) (or U(3)) which unifies the three families in
one horizontal triplet~\cite{SU3H,SU3H1}.  In this way, the
inter-family hierarchy can be related to the SU(3) breaking pattern.

In the context of SO(10)$\times$SU(3), the three families of fermions
fit into the representation $(\R{16},\R{3})=\R{16}^i$, where $i=1,2,3$
is a family SU(3) index.  The family symmetry forbids them to have
Yukawa couplings with the Higgs $\R{10}$-plet of SO(10), since this is
singlet of flavour while the bilinear of fermions transforms as
$\R{3}\times\R3=\Rb3+\R6$.  Therefore, in order to generate fermion
masses one needs a higher order operator of the form
$G_{ij}\R{16}^i\R{16}^j\R{10}$, where $G_{ij}$ is a field-dependent
effective Yukawa coupling, that transforms as $\Rb3\times\Rb3$ under
SU(3), for example it may be an (anti)sextet.  It is clear that the
pattern of fermion mass matrices will reflect the VEV structure of $G$
i.e.\ the horizontal symmetry breaking.  However, this operator should
also be SO(10) dependent, otherwise it would leave in each family an
undesirable degeneracy between up, down and charged lepton masses.
Therefore one is led to consider operators of the form
\vspace*{-1ex}
\be 
G_{ij}(X)\R{16}^i\R{16}^j\R{10}\,, 
\ee 
where $X$ is a set of SO(10) fields, and we will see that one can
efficiently take $\R{45}$-plets of SO(10). These higher order
operators can be effectively generated from renormalizable couplings
involving heavy vector-like fermions, so called \emph{universal-seesaw
mechanism}~\cite{universal}.

In this paper, we follow the idea~\cite{z,br} that the Yukawa matrices
are built as a linear combination of three fixed rank-one projectors
in flavour space, i.e.\ three $3\times3$ symmetric rank-one matrices
$\hat P_{1,2,3}$:\footnote{One may choose also non symmetric rank-one
projectors, but at the price of loosing predictivity.}
\be
\label{eq:gen} 
\hat\l_f \propto \e_f^2\, \hat P_1+\e_f\, \hat
P_2+\,\hat P_3\,, \qquad f=u,d,e,\nu_D \,.  
\ee
Only the coefficients depend on the fermion type ($f=u,d,e,\nu_D$)
while the three disoriented rank-one projectors are common to all
types.

We will build models where the matrices $\hat P_n$ are generated from
the VEVs of flavon scalar fields breaking the SU(3) symmetry, and the
factors $\e_f$ arise from the VEV of the $X$ fields at SO(10) breaking
and are related by specific Clebsch relations.

As far as the $\hat P_n$ are rank one matrices with generically order
one entries, the eigenvalues will roughly follow the
hierarchies~(\ref{eq:up}). We will see that these coefficients are
realized naturally using the VEV of a single $\R{45}_X$ field with a
generic orientation compatible with the SO(10) breaking to the
Standard Model group.  Then the breaking of flavour realized by the
$\hat P_n$ projectors not only generates the flavour mixing angles,
but also the right deviations of the eigenvalues from exact hierarchy.
Namely, there will be a link between the deviation in the D, E sectors
and the largeness of the Cabibbo angle.

We will provide the model realizations of this ansatz that also
satisfy the naturality conditions, namely the dangerous D=5 operators
inducing proton decay are automatically suppressed and the
supersymmetric flavour changing effects are under control.
These models will lead to three different cases of the
general ansatz~(\ref{eq:gen}) for the fermion mass matrices.  One of
these cases will be found in perfect agreement with the observed
masses, mixing angles and CP violation, and quite interestingly this
case points to an exact grand unification of \emph{determinants} of
the Yukawa matrices at the GUT scale:
\be
\det\,\hat\l_u=\det\,\hat\l_d=\det\,\hat\l_e=\det\,\hat\l_{\nu_D}\,.
\ee

The paper is organized as follows: in section~\ref{sec:so10} we
introduce the ``universal seesaw'' mechanism for fermion masses, and
discuss the VEV patterns for the SO(10) and SU(3) symmetry breaking.
In section~\ref{sec:realizations} we describe the model realizations
in terms of renormalizable operators and discover three interesting
cases corresponding to the Yukawa couplings unification in the first,
second or third fermion family.  Singling out the second case, in
section~\ref{sec:detunif} we will describe the unification of Yukawa
determinants and its predictions for fermion masses and the $\tb$
parameter.  In section~\ref{sec:anal} we will solve analytically the
model in terms of its parameters and also perform a numeric fit.

A preliminary short version of this work has been presented
in~\cite{proc}.

\section{Universal seesaw in SO(10) and flavour symmetry}
\label{sec:so10}

Along the lines of~\cite{z}, we describe the generation of effective
Yukawa couplings at the GUT scale via the \emph{universal seesaw}
mechanism that employs SO(10) gauge symmetry together with the SU(3)
horizontal symmetry.

The model uses Higgs fields in the following representations of
SO(10): $\R{16}$, $\Rb{16}$, $\R{10}$, $\R{45}$,
$\R{54}$.\footnote{Motivated by minimality arguments, we avoid using
huge representations like \R{210}, $\R{126}+\Rb{126}$ etc\ldots} For
later reference we recall their decomposition in terms of the standard
SU(5) and Pati-Salam SU(4)$\times$SU(2)$\times$SU(2) subgroups:
\bea
 \R{16}&=&\R1+\Rb5+\R{10}=(\R4,\R2,\R1)+(\Rb4,\R1,\R2)\,,\0\\
\Rb{16}&=&\R1+\R5+\Rb{10}=(\Rb4,\R2,\R1)+(\R4,\R1,\R2)\,,\0\\
 \R{10}&=&\R5+\Rb5=(\R6,\R1,\R1)+(\R1,\R2,\R2)\,,\0\\
 \R{45}&=&\R1+\R{10}+\Rb{10}+\R{24}=(\R{15},\R1,\R1)+(\R1,\R1,\R3)+(\R1,\R3,\R1)+(\R6,\R2,\R2)\,,\0\\
 \R{54}&=&\R{15}+\Rb{15}+\R{24}=(\R1,\R1,\R1)+(\R1,\R3,\R3)+(\R{20}',\R1,\R1)+(\R6,\R2,\R2).\0
\eea

We assume that SO(10) is broken to the Standard Model group in a
single step by means of the set of Higgs fields $\R{16}_H +
\Rb{16}_H$, $\R{54}_H$ and $\R{45}$, with VEVs of the same order. In
particular the $\R{16}_H$ and $\Rb{16}_H$, with VEV towards
$\R{1}_{\SU(5)}$, breaks SO(10) down to SU(5), while the $\R{54}$ with
VEV $\propto \mbox{diag}(1,1,1,-3/2,-3/2)$ breaks SO(10) down to
SU(4)$\times$SU(2)$\times$SU(2). The intersection of these two
breaking channels leads to the Standard Model symmetry group
SU(3)$\times$SU(2)$\times$U(1). In addition, internal consistency of
the Higgs sector requires the presence of fields in the $\R{45}$
representation.

We assume that we have three $\R{45}$-plets, $\R{45}_{BL}$,
$\R{45}_R$, $\R{45}_{X}$. The first, with direction towards the
$(15,1,1)$ component,
$\langle\R{45}_{BL}\rangle\propto(1,1,1,0,0)\otimes \sigma$, can be
used for the solution of the doublet-triplet splitting problem via the
Missing VEV mechanism~\cite{dimwilc}. A second one, with VEV
orthogonal to that of $\R{45}_{BL}$, towards the $(1,1,3)$ component:
$\langle\R{45}_R\rangle\propto(0,0,0,1,1)\otimes\sigma$.  Finally we
have a $\R{45}_X$ with generic VEV towards both these
directions.\footnote{The superpotential(s) structure realizing the VEV
pattern and a detailed analysis can be found in~\cite{proton,
potential, sextets}, and the needed form of the potential may be
motivated with an additional discrete symmetry or just rely on the
nonrenormalization theorem. The essence of the Missing VEV mechanism
also motivates the use of the three adjoint fields $\R{45}_X$,
$\R{45}_R$, $\R{45}_{BL}$.}

The Higgs doublets $(h_u,h_d)=\phi$, which induces the electroweak
breaking and the fermion masses is sitting in the $(1,2,2)$ component
of the $\R{10}$-plet. The $(6,1,1)$ component contains the colored
higgses $T$, $\overline T$, that should be very heavy, of the order of
the GUT scale, in order to suppress proton decay via dimension 5
operators~\cite{dimfive}.

As far as the breaking of the flavour SU(3) symmetry is concerned, we
will use three sextets $\chi^{ij}_{1,2,3}$ with disoriented rank-one
VEVs~\cite{anselm, disoriented}. These VEVs generically can be
presented as $\langle\hat\chi_n\rangle=U_n\,{\rm diag}(0,0,1)\,U^t_n$,
with $U_n$ being 3$\times$3 unitary matrices. Alternatively a sextet
field $\chi^{ij}_n$ may be built as effective tensor product of
triplets $\chi^{ij}=\xi^i\cdot\xi^j$~\cite{SU3H2}. If one introduces
three triplets $\xi_n^i$ ($n=1,2,3$) having misaligned VEVs in SU(3)
space, then the tensor products $\xi_n^i\cdot\xi_n^j$ automatically
provide effective $\chi^{ij}_n$ with VEVs of rank-one.

\subsection{Universal seesaw}

We introduce heavy vector-like fermions $\R{16}'_i$, $\Rb{16}'_i$ that
are antitriplets of flavour and consider the following superpotential
terms, allowed by the SO(10)$\times$SU(3) symmetry:
\be
\label{eq:pot}
W=\gamma\,\R{10}\, \R{16}^{i}\R{16}'_i \,+ \gamma'
\,\Sigma\,\R{16}^{i}\,\overline\R{16}'_i\,+\overline\R{16}'_i\,M_{\rm heavy}^{ij}\R{16}'_j
\ee
where $\R{10}$ is the fundamental Higgs and $\Sigma$ is a Higgs field
that could be in the $\R1$, $\R{45}$ or $\R{210}$ representations and
whose role will be exploited in the next section. The flavour
structure is encoded in $\hat M_{\rm heavy}$, while the first two terms
are flavour universal.  Also, $\hat M_{\rm heavy}$ is still an effective
SO(10) operator.

\begin{figure}[t]%
\centerline{\setlength{\unitlength}{6em}
\newcommand\clap[1]{\hbox to 0pt{\hss#1\hss}}
\newcommand\cclap[1]{\vbox to 0pt{\vss\hbox to 0pt{\hss#1\hss}\vss}}
  \begin{picture}(4,1.6)(.1,-0.4)
    \put(.5,.1){\clap{$\R{16}^i$}}
    \put(0,0){\line(1,0){1}}
    \put(1,0){\line(0,1){1}}
    \put(1.1,.8){$\R{10}$}
    \put(1.5,.1){\clap{$\R{16}'_i$}}
    \put(1,0){\line(1,0){1}}
    \put(2,0){\cclap{$\times$}}
    \put(2,-.2){\cclap{$M^{ij}_{\rm heavy}$}}
    \put(2,0){\line(1,0){1}}
    \put(2.5,.1){\clap{$\overline\R{16}'_j$}}
    \put(3,0){\line(0,1){1}}
    \put(3.1,.8){$\langle\Sigma\rangle$}
    \put(3,0){\line(1,0){1}}    
    \put(3.5,.1){\clap{$\R{16}^j$}}
  \end{picture}}
\vspace*{-2ex}
  \caption{Universal seesaw mechanism for Yukawa couplings.\label{fig:seesaw}}
\end{figure}
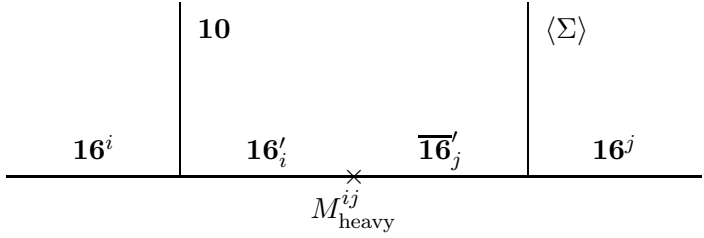

Near the GUT scale where some higgses develop a VEV, the heavy
fermions $\R{16}'$ get mass and decouple, so the light fermions
$\R{16}$ acquire an effective Yukawa-like coupling with $\R{10}$ that
is approximately given by the SO(10) ``seesaw'' formula:
\be
\label{eq:yukSO10}
W_{\rm light}=G_{ij}\,\R{10}\,\R{16}^i\,\R{16}^j\,,
           \qquad 
    \hat G=\gamma \,\gamma' \, \langle\Sigma\rangle\,\hat M_{\rm heavy}^{-1}\,.
\ee
We begin to see from here that the Yukawa couplings of the light
fermions will be proportional to the \emph{inverse} of the heavy
fermions mass matrices.

To see this happen in detail, it is useful to decompose the fermion
fields under the Pati-Salam group as
$\R{16}^i=f^i(\R4,\R2,\R1)+f^{ci}(\bar\R4,\R1,\R2)$ and
$\R{16}'_i=\F_i(\R4,\R2,\R1)+F^c_i(\bar\R4,\R1,\R2)$,
$\overline\R{16}'_i=\F^c_i (\bar\R4,\R2,\R1)+F_i(\R4,\R1,\R2)$. Note
that $f=q,l$ and $\F=q',l'$ are weak isodoublets while
$f^c=u^c,d^c,e^c,\nu^c$, $F=u',d',e',\nu'$., are weak
isosinglets. With this decomposition, we can illustrate the couplings
present in~(\ref{eq:pot}), before seesawing, as:
\vspace*{-2ex}
\be
\label{eq:fullcouplings}
\def\arraystretch{1.2}
\begin{array}{ccc}
 & {\begin{array}{ccc} \,f^c & \,\,\,\;F^c & \,\,\;\F^c
\end{array}}\\ \vspace{2mm}
\begin{array}{c}
f \\ F \\ \F  
\end{array}\!\!&{\left(\begin{array}{ccc}
    0 & {\gamma}\phi  & \mu_\F \\ \mu_F  & \hat{M}_F & 0 \\
{\gamma}\phi
 & 0 & \hat{M}_\F \end{array}\right)}
\end{array}
\vspace*{-1ex}
\ee
where $\hat M_{\rm heavy}$ is decomposed in $\hat M_F$ and $\hat M_{\cal
F}$ , respectively the mass matrices for (unmixed) isosinglets and
isodoublets (recall that heavy fermions $F$-$F^c$ and $\cal F$-${\cal
F}^c$ have vector-like masses) and $\mu_F$, $\mu_\F$ are the
projections of $\gamma'\langle\Sigma\rangle$ on the isosinglets and
isodoublets channels. Recall that the latter entries are flavour blind
and the nontrivial flavour structure is contained only in the matrices
$\hat M_F$ and $\hat M_{\cal F}$.

This form makes explicit the seesaw mechanism that follows from the
mixing between the isosinglets $f^c$ and $F^c$ and between the
isodoublets $f$ and $\cal F$.  After this mixing one ends up with an
effective Yukawa couplings of the light fermions $f$, $f^c$ to the
Higgs $\phi$, as in the Standard Model: $\phi\, f\hat \l_{f^c}f^c$.
Decomposing further the SU(4) of Pati-Salam in quark and lepton
channels one recovers the Standard-Model Yukawa
couplings~(\ref{eq:SMyuk}) with Yukawa matrices given
by:\footnote{Without the risk of confusing notations we denote here
with $f$ the fermion type, instead of the light isodoublets.}
\be
\label{eq:fullseesaw}
\hat\l_{f}=\gamma \,\mu_F \,\hat M_F^{-1}+\gamma \,\mu_\F\, \hat
M_\F^{-1}\,,
\rlap{$\qquad\qquad f=u, d, e, \nu_D\,.$}
\ee
We see that the Yukawa couplings receive two contributions, from the
inverses of both $\hat M_F^{-1}$ and $\hat M_\F^{-1}$.  Therefore each
fermion type $u$, $d$, $\nu$, $e$ will receive one contribution from
the mass matrix of its relative heavy isosinglets
$F=u'$,~$d'$,~$\nu'$,~$e'$, and an other from that of the appropriate
heavy isodoublets $\F=q'$ or~$l'$.  We will expand below the form of
$\hat M_F$ and $\hat M_\F$ to describe explicitly the realization of
the charged fermions mass matrices.  Before doing this, we complete
this section addressing the generation of neutrino masses.

The Yukawa couplings obtained above generate Dirac masses for charged
fermions as well as for neutrinos and, as is frequent in unified
models, these neutrino masses will be unrealistically high.  Therefore
one is led to consider also Majorana mass terms for the RH neutrinos,
and to generate the LH neutrino masses by canonical seesaw.  After all
RH neutrino are neutral particles and there is no reason to forbid
their Majorana mass term.

\begin{figure}[t]%
\centerline{\setlength{\unitlength}{6em}    
\newcommand\clap[1]{\hbox to 0pt{\hss#1\hss}}
\newcommand\cclap[1]{\vbox to 0pt{\vss\hbox to 0pt{\hss#1\hss}\vss}}
  \begin{picture}(4,1.6)(.1,-0.4)
    \put(.5,.1){\clap{$\R{16}^i=\nu^{ci}$}}
    \put(0,0){\line(1,0){1}}
    \put(1,0){\line(0,1){1}}
    \put(1.1,.8){$\langle\overline\R{16}_H\rangle$}
    \put(1.5,.1){\clap{$\R{1}_i^N$}}
    \put(1,0){\line(1,0){1}}
    \put(2,0){\cclap{$\times$}}
    \put(2,-.2){\cclap{$M^{ij}_{N}$}}
    \put(2,0){\line(1,0){1}}
    \put(2.5,.1){\clap{$\R{1}_j^N$}}
    \put(3,0){\line(0,1){1}}
    \put(3.1,.8){$\langle\overline\R{16}_H\rangle$}
    \put(3,0){\line(1,0){1}}
    \put(3.5,.1){\clap{$\R{16}^j=\nu^{cj}$}}
  \end{picture}}
\vspace*{-2ex}
  \caption{Universal seesaw mechanism for RH neutrino mass..\label{fig:RHseesaw}}
\end{figure}
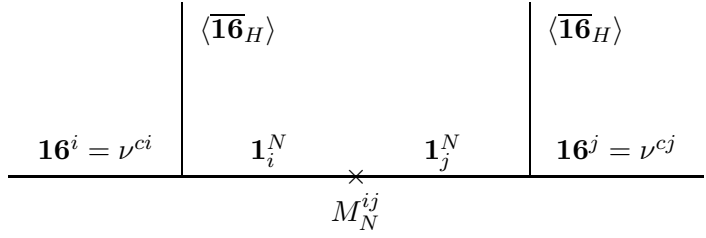

The Majorana mass for the RH neutrinos can be generated in a similar
way as above, via an \emph{universal seesaw} mechanism using the
$\Rb{16}_H$ Higgs field, whose VEV selects the RH neutrino $\nu_R$
from the $\R{16}$-plets. The relevant part of the superpotential is in
this case:
\be
\label{eq:potN}
W_N=\gamma_N\Rb{16}_H\, \R{16}^{i}\R{1}_i \,+M_{N}^{ij}\,\R{1}_i\R{1}_j
\ee
where $\Rb{16}_H$ is the Higgs whose VEV preserves SU(5) as described
above, and where we have introduced a flavour triplet of SO(10)
singlet fermions $(\R1,\Rb3)=\R{1}_i$.\footnote{Alternatively one
  could use e.g.\ a $(\R{45},\Rb3)$ multiplet.} The
flavour structure is again encoded in the mass matrix $\hat M_{N}$ of
the heavy singlets, while the first term is flavour universal.  Also,
$\hat M_{N}$ is again an effective operator that should of course be
an SO(10) singlet. We do not want at this point to dwell on the
details of its construction, that will follow lines similar to the
case of the charged fermions, described in the next section.

When the heavy singlets $\R1_i$ get mass and the $\R{16}_H$ develops a
VEV, the RH neutrino mass matrix emerges from the universal seesaw, as
depicted in figure~\ref{fig:RHseesaw}, and is given by
\be
\label{eq:MR}
\hat M_{\nu_R}=C^2 \hat M_{N}^{-1}\,,
\ee
where $C=\gamma_N\langle\Rb{16}_H\rangle$.  Then, the light (LH)
neutrino mass matrix results from the canonical seesaw with the
neutrino Dirac couplings given by~(\ref{eq:fullseesaw}), as
\be
\label{eq:doubleseesaw}
\hat M_{\nu_L} = \hat M_{\nu_D} \hat M_{\nu_R}^{-1} \hat M^t_{\nu_D}=
\frac{1}{C^2}\hat M_{\nu_D} \hat M_{N} \hat M^t_{\nu_D}\,.  
\ee
with $\hat M_{\nu_D}=v_u\hat\l_{\nu_D}$ and where $v_u$ is the ``up''
electroweak VEV, $v_u=\langle h_u\rangle=v\sin\b\simeq
174\,\GeV\,\sin\b$ in the MSSM.

We can at this point comment on the scales required for this neutrino
mass generation, that can be inferred from
relation~(\ref{eq:doubleseesaw}). Since the Dirac neutrino masses are
unified with the other fermions, they will fall in the $\GeV$ range;
in particular in these models $M_{\nu_D}\sim0.01\mbox{--}10\,\GeV$
(see later). Therefore in order to obtain realistic neutrino masses of
magnitude $\lesssim\!0.1\,\eV$ one requires the two scales $C$, $M_N$
to be related as $(C/\GeV)^2\simeq 10^{12} (M_N/\GeV)$. This leads for
example, for $M_N\sim10^{18}\,\GeV$, to $C\sim10^{15}\,\GeV$.  Since
$C=\gamma_N\langle\R{16}_H\rangle$, this requires $\gamma_N$ to be
just of order $10^{-1}$, in order to keep $\langle\R{16}_H\rangle$ at
the GUT scale.  On the other hand there is no problem in having such a
high $M_N$ decoupling scale, since the $\R{1}_i$ fermions are SO(10)
singlets and they do not interfere with the gauge group breaking.

Finally, for later convenience, we observe that introducing
dimensionless matrices also for the left and right handed neutrinos
$\hat\l_{\nu_L,\nu_R}=\hat M_{\nu_L,\nu_R}/v_u$ and for the heavy
singlets $\hat\l_N=\hat M_N \,v_u/C^2$, we can write the above
expressions in the form:
\be
\label{eq:canonical}
\hat \l_{\nu_L}=\hat \l_{\nu_D}\hat\l_{\nu_R}^{-1}\hat\l_{\nu_D}^t=\hat \l_{\nu_D}\hat\l_N\hat\l_{\nu_D}^t\,.  
\ee

\subsection{Flavour structure from rank-one decomposition}

Let us discuss now the form of $\hat M_{\rm heavy}$. Under flavour SU(3)
symmetry, $\hat M_{\rm heavy}$ transforms as $\R3\times\R3$, therefore it
can be generated via the VEVs of some flavon fields in SU(3) sextet
representation. 

Our central assumption now is that $M_{\rm heavy}$ is built as a linear
combination, with hierarchic coefficients, of three rank-one matrices
in flavour space.  The sum will give a non degenerate matrix, with
hierarchic eigenvalues.

The three rank-one matrices may for example be built out of sextet or
triplets scalar fields. For example there may be three sextets, that
break the flavour symmetry by developing a VEV, and their VEVs should
be of rank-one.  This can be achieved through some mechanism like
those described in~\cite{disoriented}.

The hierarchic coefficients will be generated as effective operators
at SO(10) breaking, and we will focus on them later; at this point it
is useful to denote them with $\a_{{\rm heavy},n}$, with $n=1,2,3$: they are
singlets under flavour but are functions of SO(10) fields $X$ and
generically transform as \R{1}, \R{45} or \R{210}. 

Summing up, we introduce three sextets $\chi^{ij}$ and mix them with
the $\a_{\rm heavy}$'s:
\be
  M_{\rm heavy}^{ij} \ = \
\sum_{n=1,2,3} \a_{{\rm heavy},n}(X)\,\chi_n^{ij}\,.
\ee
Even the heavy singlets mass matrix $M_N^{ij}$ that gives Majorana
mass to neutrinos transforms as a sextet of flavour, and thus it may
be naturally built using the same sextets $\chi_n^{ij}$, however with
different coefficients~$\a_{N,n}$:
\vspace*{-1ex}
\be
M_N^{ij} \ = \  
\sum_{n=1,2,3}\a_{N,n}\,\chi_n^{ij}\,.
\ee

At the GUT scale, where SO(10) is broken to the SM group, the
coefficients $\a_{{\rm heavy},n}$ decompose in the weak isosinglets and
isodoublets channels, and consequently $M_{\rm heavy}$ decomposes in $\hat
M_F$ and $\hat M_\F$. We can write, for all the heavy fermions
\bea
\hat M_F&=&M \big[\a_{F,1}\,\hat Q_1+\a_{F,2}\,\hat Q_2+\a_{F,3}\,\hat
  Q_3\big]\,,
\nonumber\\[.7ex]
\hat M_\F&=&M \big[\a_{\F,1}\,\hat Q_1+\a_{{\cal
  F},1}\,\hat Q_2+\a_{\F,1}\,\hat Q_3\big]\,,
\nonumber\\[0.7ex]
\hat M_N &= & M
\big[\a_{N,1}\,\hat Q_1+\a_{N,2}\,\hat Q_2+\a_{N,3}\,\hat Q_3\big]\,,
\label{eq:MF}
\eea
where the $\a$'s have been projected on the isosinglets or isodoublets
sectors and are now pure numbers, and where we have parametrized the
rank-one VEV of the sextets $\langle\chi_n\rangle$ with $M \hat
Q_n$. $M$ is the scale of flavour breaking and $Q_n$ are dimensionless
\emph{flavour projectors}.

To find the Yukawa couplings and the RH neutrino masses, given by the
inverses of these heavy mass matrices via~(\ref{eq:fullseesaw})
and~(\ref{eq:MR}), we can exploit a useful property of every
combination of rank-one projectors, namely that its inverse is also a
combination, with inverse coefficients, of (new) rank-one
projectors. Explicitly,
\be\label{eq:inversion}
     \big(\a_1\hat Q_1+\a_2\hat Q_2+\a_3\hat
     Q_3\big)^{-1}=\left(\frac1{\a_1}\hat P_1+\frac1{\a_2}\hat
     P_2+\frac1{\a_3}\hat P_3\right),
\ee
where the $\hat P_n$ are the ``reciprocal'' of the $\hat Q_n$
projectors: if one parametrizes the $\hat Q_n$ with the symmetric
product of generic complex vectors $q_n$ (flavour triplets) as $\hat
Q_n=q_nq_n^t=q_n\otimes q_n$, the $\hat P_n$ can parametrized with
three vectors $p_n$, that are in fact the reciprocal of the $q_n$
ones: $p_n=\frac12\e_{nmr}q_m\wedge q_r/D_q$, with
$D_q=q_1\cdot(q_2\wedge q_3)$. In this notation, the inverse heavy
mass matrices are then:
\bea
\hat M_F^{-1}&=&M^{-1}\big[\a^{-1}_{F,1}\,\hat P_1+\a^{-1}_{F,2}\,\hat P_2+\a^{-1}_{F,3}\,\hat
  P_3\big]\,,
\nonumber
\\[.7ex]
\hat M_\F^{-1}&=&M^{-1}\big[\a^{-1}_{\F,1}\,\hat P_1+\a^{-1}_{{\cal
  F},1}\,\hat P_2+\a^{-1}_{\F,1}\,\hat P_3\big]\,,
\nonumber
\\[0.7ex]
\hat M_N^{-1} &=&M^{-1}
\big[\a^{-1}_{N,1}\,\hat P_1+\a^{-1}_{N,2}\,\hat P_2+\a^{-1}_{N,3}\,\hat P_3\big]\,.
\eea

From~(\ref{eq:fullseesaw}),~(\ref{eq:MR}) we then find the form of
Yukawa couplings and RH neutrino masses, reaching the conclusion that
they are all written as combinations of the same three rank-one
flavour projectors~$\hat P_n$:
\bea
\hat\l_f&\propto&  \Big[\a_{f,1}\,\hat P_1+\a_{f,2}\, \hat P_2+\a_{f,3}\,\hat
  P_3\Big]\,,\qquad f=u,d,e,\nu_D
\\[.7ex]
\hat \l_{\nu_R}&\propto& \Big[\a_{\nu_R,1}\,\hat P_1+\a_{\nu_R,2}\, \hat
  P_2+\a_{\nu_R,3}\,\hat P_3\Big]\,,
\eea
with coefficients that are
\bea
\a_{f,n}&=&\frac{\mu_F}{\a_{F,n}}+\frac{\mu_\F}{\a_{\F,n}}\,,\\
\a_{\nu_R,n}&=&\frac{C^2}{\a_{N,n}}\,.
\eea

As far as the light (LH) neutrino Majorana mass matrix is concerned,
we finally have to perform the canonical seesaw~(\ref{eq:canonical}),
and we can exploit an other interesting property of rank-one
combinations. Indeed, the following ``seesaw'' relation holds:
$$
\big(\a_{1}\, \hat P_1+\a_{2}\, \hat P_2+\a_{3}\,\hat P_3\big)\big(\b_{1}\, \hat P_1+\b_{2}\,
  \hat P_2+\b_{3}\,\hat P_3\big)^{-1}\big(\a_{1}\, \hat P_1+\a_{2}\, \hat P_2+\a_{3}\,\hat P_3\big)=
$$
\vspace*{-1.1em}
\be
=\left(\frac{\a_{1}^2}{\b_1}\, \hat P_1+\frac{\a_{2}^2}{\b_2}\, \hat P_2+\frac{\a_{3}^2}{\b_3}\,\hat P_3\right),
\label{eq:seerel}
\ee 
i.e.\ the seesaw acts on the coefficients only, but does not change
the flavour projectors $\hat P_n$.  Hence, also the neutrino mass
matrix is built as a combination of the same flavour projectors:
\be
\hat\l_{\nu_L}=\hat\l_{\nu_D}\hat\l_{\nu_R}^{-1}\hat\l_{\nu_D}^t
\propto 
\Big[\a_{\nu_L,1}\,\hat P_1+\a_{\nu_L,2} \,\hat P_2+\a_{\nu_L,3}\,\hat P_3\Big]\,,\qquad
\a_{\nu_L}=\frac{\a_{\nu_D}^2}{\a_{\nu_R}}\,.
\label{eq:seeneu}
\ee

We have thus shown that the basic idea of dealing with combinations of
common rank-one projectors is not spoiled by the seesaw generation of
neutrino masses, and this will allow below the model to be nicely
predictive in this sector.

\subsection{Choice of $\R{45}_R$ in the ``right'' direction}

In the universal seesaw for Yukawa couplings, one can take the Higgs
field $\Sigma$ in the $\R{45}$ representation. Then, the possible VEV
directions are (15,1,1) and (1,1,3) of Pati-Salam. Choosing the
(1,1,3) direction leads to a number of interesting consequences, that
we will briefly describe. Let's denote $\Sigma$ in this case as
$\R{45}_R$.  When its VEV is chosen in that direction, i.e.\
$\gamma'\langle\R{45}_R\rangle=\mu_F(1,1,3)$, we have that it couples
only light and heavy isosinglets, but not isodoublets.
Correspondingly $\mu_\F=0$ and in the couplings matrix~(\ref{eq:fullcouplings}) the $f$-$\F^c$ entry is zero:
\vspace*{-1ex}
\be
\def\arraystretch{1.2}
\begin{array}{ccc}
 & {\begin{array}{ccc} \,f^c & \,\,\,\;F^c & \,\;\F^c
\end{array}}\\ \vspace{2mm}
\begin{array}{c}
f \\ F \\ \F   
\end{array}\!\!&{\left(\begin{array}{ccc}
    0 & {\gamma}\phi  & 0 \\ \mu_F  & \hat{M}_F & 0 \\
{\gamma}\phi
 & 0 & \hat{M}_\F \end{array}\right)}
\end{array}
\vspace*{-1ex}
\ee
The first consequence to be noted is that now only the $f^c$, $F^c$
participate in the seesaw, and instead of~(\ref{eq:fullseesaw}), it
yields a simpler expression for the Yukawa matrices:
\be
\label{eq:lambdaseesaw}
\hat\l_f=\gamma \,\mu_F \,\hat M_F^{-1}\,.  
\ee 
This expression has the property that a given hierarchy of the heavy
isosinglets fermions $F$ is directly reproduced in the light Yukawa
couplings, while in the general case the light Yukawa was given by the
sum of the inverses $\hat M_F^{-1}$, $\hat M_\F^{-1}$. Each ``light''
fermion type $f$ receives now a Yukawa coupling only from the
``heavy'' relative sector of isosinglets $F$, proportional to their
inverse mass matrix in flavour space.  This is a particularly
desirable situation, since every fermion sector follows an almost
exactly hierarchic pattern, that is easier to obtain from a single
inverse matrix, rather than the sum of two.  In this scenario
therefore, we observe that the heavy $F$ fermions should exhibit an
\emph{inverted hierarchy} pattern, with the lightest particle being
the heavy correspondent of the ``top''.

In addition to these facts, we note also two other remarkable features
of the universal seesaw using a $\R{45}_R$: 1) the LLLL (dominant)
part of the D=5 proton decay is automatically eliminated, because
(\R1,\R1,\R3) does not couple the light and heavy
isodoublets~\cite{z,proton}; 2) the sfermion mass matrices are
automatically aligned with the square of the fermion ones, thus
avoiding the SUSY flavour problem (see e.g.~\cite{susyflavour}).
We conclude that the breaking in the SU(2)$_R$ direction automatically
allows the correct mass generation, suppresses the proton decay and
avoids the SUSY flavour problems.

\subsection{Flavour Clebsches from $\R{45}_X$}

We focus now on the SO(10) structure for the construction of the
hierarchic coefficients $\a_{F,1,2,3}$. We will argue that the use of
a $\R{45}_X$ Higgs field alone predicts hierarchy parameters that are
in agreement with the phenomenologically observed pattern.

We first observe that since the known fermion masses follow a direct
hierarchy, one needs for the heavy fermions an inverted pattern, i.e.\
$\a_{F,1,2,3}\sim1,\e,\e^2$.  We may generate small parameters from
the ratio of a VEV $\langle\R{45}_X\rangle$ to some higher scale $V$,
and the sequence can be generated by taking successive powers. For
example the operator form of $M_{\rm heavy}$ may be:
\be
\label{eq:Mheavy-ex}
M_{\rm heavy}^{ij} \propto 
\Big[\chi^{ij}_1 +\frac{\R{45}_X}{V}\chi^{ij}_2
    +\frac{\R{45}_X^{\!2}}{V^2}\chi^{ij}_3\Big]\,,
\ee
whose realization as effective operator will be addressed in detail in
the next section. 

At SO(10) breaking, when $\R{45}_X$ develops a VEV, it is clear that
the three coefficients will be hierarchic according to the small ratio
$\langle\R{45}_X\rangle/V$, and this, projected on the different
components of the heavy $\R{16}$-plets, will give rise to different
small parameters. Let's denote them $\e_u$, $\e_d$, $\e_e$,
$\e_{\nu_D}$ for isosinglets and $\e_q$, $\e_l$ for isodoublets. When
$M_{\rm heavy}$ is decomposed in $M_F$, $M_\F$ and all fields get their
VEV, we have:
\bea
\hat M_F&\propto& \big[\hat Q_1+\e_F\hat Q_2+\e_F^2\hat Q_3\big]\,,\qquad F=u,d,e,\nu_D\\[0.7ex]
\hat M_\F&\propto& \big[\hat Q_1+\e_\F \hat Q_2+\e_\F^2\hat Q_3\big]\,,\qquad \F=q,l\,.
\eea
The Yukawa couplings will be proportional to the inverse of the
isosinglets matrix $\hat M_F$ (see~(\ref{eq:lambdaseesaw})) leading
to:
\vspace*{-1ex}
\be
\label{eq:lf3}
  \hat\l_f \propto\Big[\e_f^2\,\hat P_1\,+\e_f\,\hat P_2\,+\hat P_3\Big]\,,
  \qquad f=u,d,e,\nu_D\,,
\ee
that inherits the small hierarchy parameters from $M_F$, but in
inverted order.

\pagebreak[3]

The $\e_f$ parameters are determined by the breaking pattern of
$\R{45}_X$, so an important question is: can they be realistic?  The
answer to this question is not only that they can, but also that this
fact is a prediction when one uses the $\R{45}_X$.  To see this, let
us parametrize the generic allowed direction of
$\langle\R{45}_X\rangle$ as $\langle\R{45}_X\rangle/V =
\e_{15}(15,1,1)\, +\e_3(1,1,3)$. Then the different small parameters
$\e_u$ $\e_d$, $\e_e$, $\e_{\nu_D}$, $\e_q$, $\e_l$ are all determined
by the two $\e_{15}$, $\e_3$ with some SO(10) Clebsch-Gordan factors,
as follows:
\be
\begin{array}{lll}
\e_u=\e_{15}+\e_3  \qquad & \e_\nu=-3\e_{15}+\e_3\\
\e_d=\e_{15}-\e_3         & \e_e=-3\e_{15}-\e_3\\[1ex]
\e_q=-\e_{15}              &\e_l=3\e_{15}\,.
\end{array}
\ee 
Hence, since they depend on just two parameters, they are not
independent, but satisfy the following relations:
\be
\e_e=-\e_d-2\e_u\,,\qquad\e_{\nu_D}=-2\e_d-\e_u\,,\qquad\e_l=-3\e_q=\frac32(\e_d+\e_u)\,.
\label{eq:soten}
\ee

The first relation is particularly interesting, since it leads to the
following two observations. Firstly, once one ensures that $|\e_u|
\ll|\e_d|$, the equality of hierarchies in the D, E sectors is
predicted: $|\e_e|\simeq|\e_d|$.  Therefore the approximate D-E
symmetry appears as a prediction, assuming just that the U sector is
more hierarchic.  Secondly, still from the first relation, we observe
that the signs of $\e_e$, $\e_d$ are nearly opposite, and this will be
crucial for explaining the deviations from exact hierarchies in the E
and D sectors, in section~\ref{sec:anal}.  Therefore we conclude that
that the use of a $\R{45}_X$ with VEV in a generic direction
compatible with SO(10) breaking to SM naturally accommodates the
pattern of hierarchies observed in nature.

Regarding the neutrino sector, one similarly introduces an operator
for the singlet states $M_N^{ij}$ built with the \emph{same} flavon
sextets, that when inverted generates a RH neutrino mass matrix in the
form of a combination of the same $\hat P_n$ projectors, with a
different hierarchic parameter $\e_{\nu_R}$.  

The only important difference with respect to the case of Yukawa
couplings is that the three coupling constants the were hidden in the
previous expressions (like for example~(\ref{eq:Mheavy-ex})) now play
a role: indeed there one was able to reabsorb the coupling constants
in the normalization of the $\chi_n$ sextet fields, but this can be
done only once.  Now one can only absorb one in the overall scale and
an other in the hierarchic parameter $\e_{\nu_R}$, while the third
will remain as an explicit parameter, $\a$:
\be
  \hat\l_{\nu_R} \propto\Big[\e_{\nu_R}^2\,\hat P_1\,+\e_{\nu_R}\,\hat P_2\,+\frac1\a\hat P_3\Big]\,.
\ee
This additional constant will appear also in the mass matrix of the
light neutrinos: in fact using $\hat\l_{\nu_R}$ in the seesaw with the
Dirac neutrino matrix~(\ref{eq:lf3}) and exploiting for the rank-1
flavour decomposition the seesaw
relation~(\ref{eq:seerel}),~(\ref{eq:seeneu}), we find the form of the
LH neutrino mass matrix:
\be
\label{eq:Lnu}
\hat\l_{\nu_L} \propto\Big[\e_{\nu_L}^2\,\hat P_1\,+\e_{\nu_L}\,\hat P_2\,+\a\hat P_3\Big]\,,\qquad\mbox{with~}\e_{\nu_L}=\frac{\e_{\nu_D}^2}{\e_{\nu_R}}\,.
\ee

We have found that the light neutrino mass matrix finally has the same
form of combination of the three $\hat P_n$ rank-1 projectors, but
with non-hierarchic coefficients. The complex numbers $\e_{\nu_L}$ and
$\a$ will be the two free parameters to be used when testing the model
predictions in the neutrino sector.

\section{Three Yukawa unification cases}
\label{sec:threecases}

Coming back to Yukawa matrices, the combination~(\ref{eq:lf3}) has
$\a_{f,1,2,3}=\e_f^2:\e_f:1$, and one can observe that, neglecting the
corrections of order $\e_f$, the fermions of the third generation will
roughly have all the same Yukawa coupling, corresponding to the
eigenvalue of $\hat P_3$. This realizes the traditional $t$-$b$-$\tau$
unification scheme, where at GUT scale the Yukawa of the third
generation get unified. In the MSSM this requires the parameter
$\tb=v_u/v_d$ to have a very large value, $\simeq55$.

However, we note that the pattern of known fermion masses does not
require this to happen, it just requires the three coefficients to be
hierarchic within each fermion sector. Therefore one may add an
overall power of $\e_f$ to the construction of the last section and
look for Yukawa couplings of the form:
\be
\label{eq:Lgen}
  \hat\l_f=\l\,\e_f^{n-2}\,\Big[\e_f^2\,\hat P_1\,+\e_f\,\hat P_2\,+\hat P_3\Big]
  \qquad\qquad f=u,d,e,\nu_D
\ee
where the power of $\e_f$ that we have added is a generic integer
because it should be generated as effective powers of $\R{45}_X/V$.

\begin{figure}[t]%
\centerline{\epsfig{file=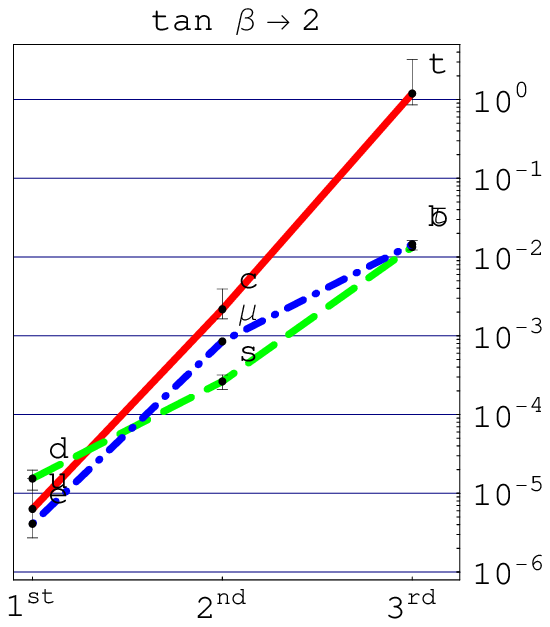,height=25em,width=.33\textwidth}%
  \epsfig{file=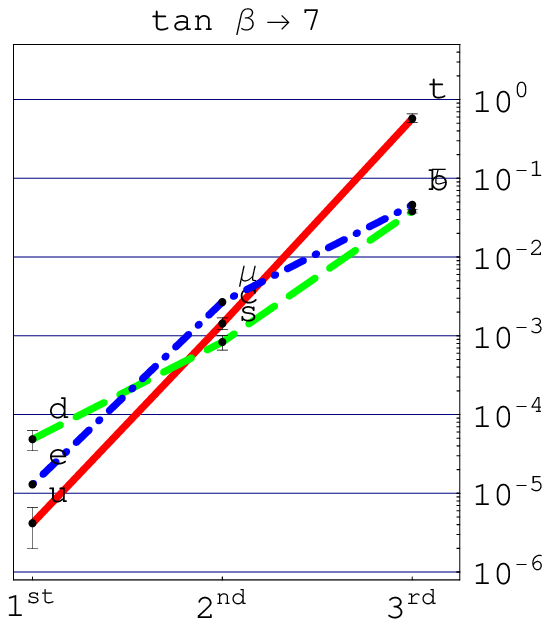,height=25em,width=.33\textwidth}%
  \epsfig{file=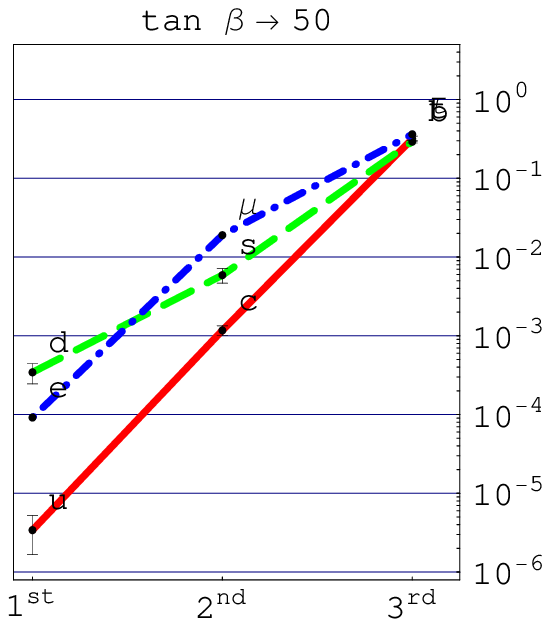,height=25em,width=.33\textwidth}}%
  \caption{Yukawa couplings at GUT scale calculated from experimental
  charged fermions masses, assuming MSSM running and including
  uncertainty from experiments and from $\l_t$ renormalization. U, D, E are
  respectively red (continuous), green (dashed) and blue
  (dot-dashed).\label{fig:abc}\vspace*{2em}}
\end{figure}

This form, for $n=0$,$1$,$2$, singles out three interesting cases of
approximate Yukawa unifications, that corresponds in the MSSM to the
cases illustrated in figure~\ref{fig:abc}, for different values of $\tb$:
\begin{itemize}  
\item[I)] for $n=0$, the coefficients are $1,\e_f^{-1},\e_f^{-2}$, and
  the GUT Yukawa couplings of the \emph{first} generation are
  approximately unified. In the MSSM this requires a very low value of 
  $\tb$~($<2$);
\item[II)] for $n=1$, the coefficients are\ $\e_f,1,\e_f^{-1}$, we find
  an approximate unification of the \emph{second} generation, at
  moderate $\tb$ ($\simeq10$);
\item[III)] for $n=2$, the coefficients are\ $\e_f^2,\e_f,1$, and large
  $\tb$ ($>50$), one reaches unification of the \emph{third}
  generation, as in the known case of $t$-$b$-$\tau$ unification.
\end{itemize}
In figure~\ref{fig:abc} we have plotted the Yukawa couplings
calculated from experimentally known masses, assuming MSSM
renormalization, and including experimental unecrtainties as well as
uncertainty from large $\l_t$ renormalization.

One may observe that in the first two cases unification of the first
or second generation Yukawa couplings is not exact (there is a
splitting of a factor of 4 in the ratios $\l_d/\l_e$ or
$\l_s/\l_\mu$); This is due to yukawa eigenvalues deviating from exact
hierarchy (a straight line) and we will see that this deviation is
explained, in the present framework, as a consequence of the magnitude
of the Cabibbo mixing. We will also show that among the three cases
the second one (II) gives optimal agreement with the data. In this
case, also, the intermediate $\tb$ is in the right range, favored by
the MSSM Higgs sector constraints~\cite{higgs}.

\pagebreak[3]

\section{SO(10) realizations}
\label{sec:realizations}

To realize these three models of Yukawa couplings, for generic $n$,
one should look for effective heavy fermion mass operators of the
form,
\be
\label{eq:MgenOP}
M_{\rm heavy}^{ij} \propto 
\bigg(\frac {\R1_X}{\R{45}_X}\bigg)^{\!n} \,
\bigg[\chi^{ij}_1\, +\frac{\R{45}_X}{{\R1_X}}\,\chi^{ij}_2
    \,+\frac{\R{45}_X^{\!2}}{{\R1_X}^2}\,\chi^{ij}_3\bigg]\,,
\ee
where again the $\chi^{ij}_{1,2,3}$ are sextets of flavour with VEV
that gives the $\hat Q_{1,2,3}$ projectors, and $\R1_X$ is a singlet
of SO(10)$\times$SU(3) with VEV larger than $M_{\rm GUT}$.  Equivalently
one may try to directly generate the light Yukawa couplings as
effective operators of the form
\be
\label{eq:LgenOP}
\l_{ij} \propto 
\frac{\R{45}_R}{{\R1_X}^2}\bigg(\frac{\R{45}_X}{{\R1_X}}\bigg)^{\!n-2} \,
\bigg[\frac{\R{45}_X^{\!2}}{{\R1_X}^2}\,\xi_{ij}^1\, +\frac{\R{45}_X}{{\R1_X}}\,\xi_{ij}^2
    \,+\xi_{ij}^3\bigg]\,,
\ee
where the $\xi^{1,2,3}_{ij}$ are now antisextets of flavour and their
VEV are the $\hat P_{1,2,3}$ projectors.

All these effective operators can be realized using just
renormalizable interactions introducing additional fields, that vary
for the different cases. In general one has to add additional fermion
multiplets beyond the $\R{16}'$ $\Rb{16}'$, that can be taken to be
vector-like as well; they will be denoted as $\R{16}''$, $\Rb{16}''$,
$\R{16}'''$, $\Rb{16}'''$, etc. We then allow couplings of all the
vector-like fermions together, but for simplicity we couple the light
multiplet $\R{16}$ just with $\R{16}'$, $\Rb{16}'$, to realize the
universal seesaw. The coupling matrix can be illustrated as:
$$
\def\lline{\!\vline\hfill}%
\def\arraystretch{1.2}
 \ba[t]{l|cccc|}
 &\R{16}^i&\R{16}'_i&\R{16}^{\prime\prime}&\ldots \\
 \hline
 \R{16}^j &   &\R{10}&&\\
 \cline{3-5}
 \overline\R{16}'_j&\R{45}_R&\lline   &&\\
 \overline\R{16}^{\prime\prime}&   &\lline& M_{f\!ull}&\\
 \vdots&   &\lline&&\\
 \hline
 \ea
$$ and $M_{f\!ull}$ will contain all the renormalizable couplings with
SO(10) or flavour fields such as $\R{45}$, $\chi_{ij}$, that all
develop a VEV around the GUT scale.

The condition to use the universal seesaw approximation is that the
involved eigenvalues of $M_{f\!ull}$ are heavier than
$\langle\R{45}_R\rangle$; once this is assured, the mass matrix $\hat
M_{\rm heavy}$ of the heavy states is effectively given by the projection
of $M_{f\!ull}^{-1}$ into the $3\times3$ subspace of the $\R{16}'_i$,
$\Rb{16}'_j$:
\be
    \hat \l\simeq\langle\R{45}_R\rangle \hat
    M_{\rm heavy}^{-1}= \langle\R{45}_R\rangle \, (\R1,0, \ldots)\,M_{f\!ull}^{-1}\raisebox{-1.6em}{$\left(\ba{c}\R1\\0\\\vdots\ea\right)$}
\ee
where \R1 stands for the identity in flavour space,
$\delta_{ij}$. This means that to describe in the general case the
resulting Yukawa couplings it is not necessary to diagonalize the
entire coupling matrix, but it is sufficient to find the first
3$\times$3 block of the inverse of $M_{f\!ull}$.

In the following we describe some of the realizations of the Yukawa
couplings~(\ref{eq:LgenOP}) for the interesting cases $n=0,1,2$,
introducing the additional fermion multiplets and the appropriate
flavon fields.

\subsection{Model I: first family unification, $n=0$}

For $n=0$ the three terms appearing in $\hat \l$ have negative powers,
0, -1, -2, of $\R{45}_X/{\R1_X}$.  These can be realized by introducing
three more vector-like fermion multiplets that are flavour triplets,
$\R{16}^{\prime\prime\,i}_{1,2,3}$ (with their SO(10) conjugates
$\Rb{16}^{\prime\prime\,i}_{1,2,3}$) and arranging the couplings as
follows:
$$
\def\arraystretch{1.3}
 \ba[t]{l|ccccc|}
 &\R{16}^i&\R{16}'_i&\R{16}^{\prime\prime\,i}_1&\R{16}^{\prime\prime\,i}_2&\R{16}^{\prime\prime\,i}_3\\
 \hline
 \R{16}^j &   &\R{10}&&&\\
 \overline\R{16}'_j &\R{45}_R&   & {\R1_X} & {\R1_X} &\R{45}_X\\
 \Rb{16}^{\prime\prime\,j}_1&   &     {\R1_X}&\xi^1_{ij}&&\\
 \Rb{16}^{\prime\prime\,j}_2&   & \R{45}_X & &\xi^2_{ij}& \\
 \Rb{16}^{\prime\prime\,j}_3&   & \R{45}_X & & & \xi^3_{ij}\\
 \hline
 \ea
$$
where $\xi^{1,2,3}$ are three flavour (anti)sextets. The effective
$\hat\l_f$ of~(\ref{eq:LgenOP}) emerges here from a \emph{double
seesaw} mechanism~\cite{doubleseesaw}, and the VEVs of
$\xi_{ij}^{1,2,3}$ directly generates the rank-one flavour projectors:
$\hat P_{1,2,3}\propto\langle\xi^{1,2,3}\rangle$.  Equivalently the
$\hat M_{\rm heavy}$ mass matrix of~(\ref{eq:MgenOP}) and the $\hat
Q_{1,2,3}$ projectors can be seen to emerge from the first 3$\times$3
block of the inverse of $M_{f\!ull}$.

Such realization is not unique, for example one may realize the same
effective operators in a more symmetric fashion with less fields, by
introducing just two vector-like triplets and antitriplets
$\R{16}''_i$, $\Rb{16}''_i$, $\R{16}^{\prime\prime i}$,
$\Rb{16}^{\prime\prime i}$ along with three sextets $\chi_n^{ij}$,
with couplings as follows:
$$
\def\arraystretch{1.3}
 \ba[t]{l|cccc|}
 &\R{16}^i&\R{16}'_i&\R{16}^{\prime\prime}_i&\R{16}^{\prime\prime i}\\
 \hline
 \R{16}^j                  &         &\R{10}              &                    &         \\
 \Rb{16}'_j                &\R{45}_R & \chi_1^{ij}        &\frac12 \chi_2^{ij} & \R{45}_X\\
 \Rb{16}''_j               &         & \frac12\chi_2^{ij} &\chi_3^{ij}         & {\R1_X}       \\
 \Rb{16}^{\prime\prime\,j} &         & \R{45}_X           & {\R1_X}                  &         \\
 \hline
 \ea
$$
In this realization the VEV of the $\chi_{1,2,3}$ sextets gives the
inverse flavour projectors that appear in $\hat M_{\rm heavy}$: i.e.\
$\hat Q_{1,2,3}\propto\langle \chi_{1,2,3} \rangle$ in~(\ref{eq:MgenOP}). Then, the final $\hat\l_f$ will still be given as a
combination of $\hat P_{1,2,3}$ as from the inversion property~(\ref{eq:inversion}).

The coupling arrangments above, and the ones that will follow, are not
the most general ones with the given fields: while some zeroes are
motivated by SU(3) flavour symmetry or by SO(10), some other choices
require a further explanation (for example the choice between
$\R{45}_X$ or ${\R1_X}$). This may be linked to the presence of some
additional symmetry (e.g.\ Z$_3$) that acts differently on the fields,
or in the intertwining of gauge and flavour symmetries in a larger
group. While these directions of investigation are very interesting,
they go beyond the scope of the present work.

\subsection{Model II: second family unification, $n=1$}
\label{sec:realizationII}

For the case $n=1$, we describe first the realization using three
triplets, that are equivalent to three sextets with rank-one VEV. One
looks for a heavy mass $\hat M_{\rm heavy}$ of the form
\be
\label{eq:MH2}
\hat M_{\rm heavy} \propto 
\bigg[\frac{{\langle\R1_X\rangle}}{\langle\R{45}_X\rangle}\hat Q_1 \,+ \hat Q_2
    \,+\frac{\langle\R{45}_X\rangle}{{\langle\R1_X\rangle}}\hat Q_3\bigg]\,,
\ee
We parametrize the three sextets $Q_{1,2,3}^{ij}$ as before via the
tensor product of three triplets: $\hat Q_n\propto q_n\, {q}_n^t$, with the
$q_n$ the VEV of three flavour triplets scalar fields. Then the
effective light Yukawa matrix can be realized by introducing four more
vector-like fermion multiplets that are \emph{flavour singlets}:
$\R{16}''_1$, $\R{16}''_2$, $\R{16}''_3$ and $\R{16}'''$ (and their
conjugates) and arranging the couplings as follows:
$$
\def\arraystretch{1.3}
 \ba[t]{l|cccccc|}
 &\R{16}^i&\R{16}'_i&\R{16}''_1&\R{16}''_2&\R{16}''_3&\R{16}'''\\
 \hline
 \R{16}^i &   &\R{10}&&&&\\
 \overline\R{16}'_j &\R{45}_R&   &q_1&q_2&q_3&0\\
 \overline\R{16}''_1&   &q_1&\R{45}_X&&&\\
 \overline\R{16}''_2&   &q_2& &{\R1_X}&&\\
 \overline\R{16}''_3&   &q_3& &&&{\R1_X}\\
 \overline\R{16}'''& &0&&&{\R1_X}&\R{45}_X\\
 \hline
 \ea
$$
By inverting in the vector-like sector the mass matrix $M_{f\!ull}$
and projecting in the $\R{16}'_i$, $\Rb{16}'_j$ space one can see
that the above form of $\hat M_{\rm heavy}$ is reproduced, inverted.
Therefore, via the universal seesaw, one obtains in the light sector
the correct Yukawa matrices:
\be
\label{eq:L2}
  \hat\l\propto \bigg[ \frac {\langle\R{45}_X\rangle} {\langle\R1_X\rangle} p_1p_1^t\,+p_2p_2^t+\frac {\langle\R1_X\rangle}{\langle\R{45}_X\rangle}\,p_3p_3^t\bigg]\qquad\qquad
  f=u,d,e,\nu_D
\ee
i.e.\ $\hat P_n\propto p_np_n^t$, with the reciprocal vectors defined
again as $p_n=\frac12\e_{mnr}q_n\wedge q_r/(q_1\cdot q_2\wedge q_3)$
(as from the inversion property~(\ref{eq:inversion})).

It is interesting to note that the way to obtain direct and inverse
powers of $\R{45}_X/{\R1_X}$ together is that the first two channels
$\R{16}''_1$, $\R{16}''_2$ are realizing a ``double'' seesaw, while
the third $\R{16}''_3$ a ``triple'' one. For we stress however that in
the full seesaw the three rank-1 contributions work together, to
reproduce the three terms in $\hat \l$.

One can also note that the scale of the ${q}_{1,2,3}$ triplets can be
raised above the $\R{45}_X$ scale and the double and triple seesaws
still work as before. In the double seesaw this can be seen by
comparing the determinant of the 2-3 block with that of the full
matrix. A similar but more involved mechanism works for the triple
seesaw. Indeed the next-to-smallest eigenvalues are, in the three
cases, ${q}_1$,~~${q}_2^2/{\R1_X}$ and $\R{45}_X$. In the third case
this is valid when ${q}_n\ge {\R1_X}>\R{45}_X$.\footnote{Considering
that $\langle\R{10}\rangle\ll
\langle\R{45}_R\rangle<\langle\R{45}_X\rangle$.}
Assuming then ${q}_n\ge {\R1_X}>\R{45}_X\simeq M_{\rm GUT}$, one can
obtain that no mass eigenvalue, apart from the light fermion masses,
drops below $M_{\rm GUT}$, so there will be no thresholds to take into
account below the GUT scale. For this to happen, the scale of flavour
breaking will be higher not only of the SO(10) breaking scale, but
also of the higher ${\R1_X}$ scale.

\pagebreak[3]

Other realizations of this model can be built by using antisextets or,
for example, two triplets $q_{1,2}$ and one sextet $\chi_3$: for this
one must introduce three vector-like fermion fields: two flavour
singlets $\R{16}''_{1,2}$ and one triplet $\R{16}^{\prime\prime\prime
i}$ (and their SO(10)$\times$SU(3) conjugates) and arranging the
couplings as follows:
$$
\def\arraystretch{1.3}
 \ba[t]{l|ccccc|}
 &\R{16}^i&\R{16}'_i&\R{16}''_1&\R{16}''_2&\R{16}^{\prime\prime\prime\,i}\\
 \hline
 \R{16}^j &   &\R{10}&&&\\
 \overline\R{16}'_j &\R{45}_R&   &q_1&q_2&\R{45}_X\\
 \overline\R{16}''_1&   &q_1&\R{45}_X&&\\
 \overline\R{16}''_2&   &q_2& &{\R1_X}&\\
 \overline\R{16}^{\prime\prime\prime}_j&   &\chi_3^{ij}& &&{\R1_X}\\
 \hline
 \ea
$$

The $\hat Q_{1,2,3}$ projectors appearing in the effective operator
will be given by $\hat Q_1\propto q_1q_1^t$, $\hat Q_2\propto q_2q_2^t$ and~$\hat
Q_3\propto \chi_3$. Then the $\hat P_{1,2,3}$ projectors entering the light
Yukawa matrices will be given again by the inversion
formula~(\ref{eq:inversion}).

\subsection{Model III: third family unification, $n=2$}

The third case corresponds to the popular but troublesome unification
of the third family quarks and leptons.  It requires very large $\tb$
that is currently disfavored, and also involves a nonlinearity also
from the renormalization evolution of the large $\l_b$ Yukawa
eigenvalue, as well as more evident corrections to the $b$
mass~\cite{bthr}.  For these reasons we will not pursue the analysis
of its phenomenological viability: we will just give an example of a
possible realization of the effective operator in this last case.

For $n=2$ we look for a realization of the following effective
operator:
\be
\label{eq:MH3}
\hat M_{\rm heavy} \propto 
\bigg[\frac{{\langle\R1_X\rangle}^{\!2}}{\langle\R{45}_X^{\!2}\rangle}q_1q_1^t \,+\frac{{\langle\R1_X\rangle}}{\langle\R{45}_X\rangle}q_2q_2^t
    \,+q_3q_3^t\bigg]\,,
\ee
for which one is led to use a ``triple'' seesaw in all the three
channels, by introducing for example six new flavour singlets
multiplets $\R{16}_{1,2,3}''$, $\R{16}_{1,2,3}'''$ (and their
vector-like conjugates), with the following couplings:
$$
\def\arraystretch{1.3}
 \ba[c]{l|cccccccc|}
 &\R{16}^i&\R{16}'_i&\R{16}''_1&\R{16}''_2&\R{16}''_3&\R{16}'''_1&\R{16}'''_2&\R{16}'''_3\\
 \hline
 \R{16}^j &   &\R{10}&&&&&&\\
 \overline\R{16}'_j &\R{45}_R&   &q_1&q_2&q_3&&&\\
 \overline\R{16}''_1&  &q_1 & & &           &\R{45}&   &    \\
 \overline\R{16}''_2&  &q_2 & & &           &      & {\R1_X} &    \\
 \overline\R{16}''_3&  &q_3 & & &           &      &   & {\R1_X}  \\
 \overline\R{16}'''_1& &    &\R{45}&      &  &  {\R1_X}   &   &   \\
 \overline\R{16}'''_2& &    &      &\R{45}&  &      & {\R1_X} &   \\
 \overline\R{16}'''_3& &    &      &      &{\R1_X} &      &   & {\R1_X} \\
 \hline
 \ea
$$

\section{Unification of determinants}
\label{sec:detunif}

Case II above, with $n=1$, is particularly interesting since it leads
to an intriguing relation between Yukawa couplings. Looking at the
determinant of the $\hat\l_f$ matrices, it turns out to
factorize~as:\footnote{One has in fact $\det(\a_1\hat P_1+\a_2\hat
P_2+\a_3\hat P_3)=\a_1\a_2\a_3D_p^2$.}
\be
\det\hat\l_f \;=\;\l^3\,D_p^2\,,
\ee
with $D_p=p_1\cdot(p_2\wedge p_3)$. This expression is independent of
$\e_f$, and this means that the determinants are unified among the
four U, D, E, $\nu_D$ sectors. In other words, at GUT scale the
eigenvalues will satisfy the following exact relation:
\be
\label{eq:yukunif}
\l_e\,\l_\mu\,\l_\tau =\l_d\,\l_s\,\l_b=\l_u\,\l_c\,\l_t=\l_{\nu_{D1}}\,\l_{\nu_{D2}}\,\l_{\nu_{D3}}\,.
\ee
The symmetry structure encoded in the generation of the Yukawa matrices, for
case II, ensures that this relation is satisfied at GUT scale
independently of the model parameters and details such as the pattern
of breaking in the flavour and gauge sectors.

Obviously to rephrase~(\ref{eq:yukunif}) as relations between fermion
masses at the weak scale one has to take into account the RG running,
and this reintroduces a dependence on the model details, the value of
$\a_3|_Z$, and mainly on the supersymmetry breaking scale.

To test the consequences of~(\ref{eq:yukunif}) on the low energy masses
we recall that, in the MSSM, fermion masses and mixing angles are
defined as follows in terms of quantities at the GUT scale:
\bea
&
\ba[b]{lllll}
m_u=\l_uv\sin\beta\, \eta_{uds} R_u B_t^3 &\quad&
m_d=\l_dv\cos\beta\, \eta_{uds} R_d &\quad&
m_e=\l_ev\cos\beta\, \eta_{e\mu}R_e
\\[1ex]
m_c=\l_cv\sin\beta\, \eta_c R_u  B_t^3  &\quad&
m_s=\l_sv\cos\beta\, \eta_{uds} R_d  &\quad&
m_\mu=\l_\mu v\cos\beta\, \eta_{e\mu}R_e 
\\[1ex]
m_t=\l_tv\sin\beta\, \eta_t R_u  B_t^6 &\quad&
m_b=\l_bv\cos\beta\, \eta_b R_d  B_t   &\quad&
m_\tau=\l_\tau v\cos\beta\, \eta_\tau R_e
\ea\\[1ex]
&
\t_{12}|_Z=\t_{12}
\,,\qquad
\t_{13}|_Z=\t_{12}\,B_t^{-1}
\,,\qquad
\t_{23}|_Z=\t_{12}\,B_t^{-1}\,,
\eea
where: the factors $R_{u,d,e}$ account for the running induced by the
MSSM gauge sector, from $M_{\rm GUT}$ to $m_t$; the factor $B_t$ accounts
for the running induced by the large $\l_t$; and $\eta_{e\mu,uds,b,c}$
complete the QCD+QED running from $m_t$ down to $2\,\GeV$ for $u$, $d$, $s$
or to the respective masses for $b$, $c$.

From~\cite{BBO},  assuming $\a_3|_Z=0.118$ and $M_{\rm SUSY}=m_t$, we find
\be
\eta_{uds}=1.74\,,\quad \eta_c=2.11\,,\quad \eta_b=1.52\,,\quad
\eta_\tau=1.04\,,\quad \eta_{e\mu}=1.06\,,\quad \eta_t=1\,.
\ee
Then, we calculate
\be
R_u=3.45\,,\qquad R_d=3.36\,,\qquad R_e=1.51\,, \qquad R_{\nu_D}=1.39\,.
\ee
The dependence on $\a_3|_Z$ of all these coefficients affects mainly
the lightest quarks via $\eta_{uds}$, $\eta_c$ (10\%), all the others
have a variation of the order of 1-3\%. These uncertainties are
however correlated.  In the $R_{u,d}$ factors there is also a stronger
dependence on the supersymmetry breaking scale.

\FIGURE{%
\epsfig{file=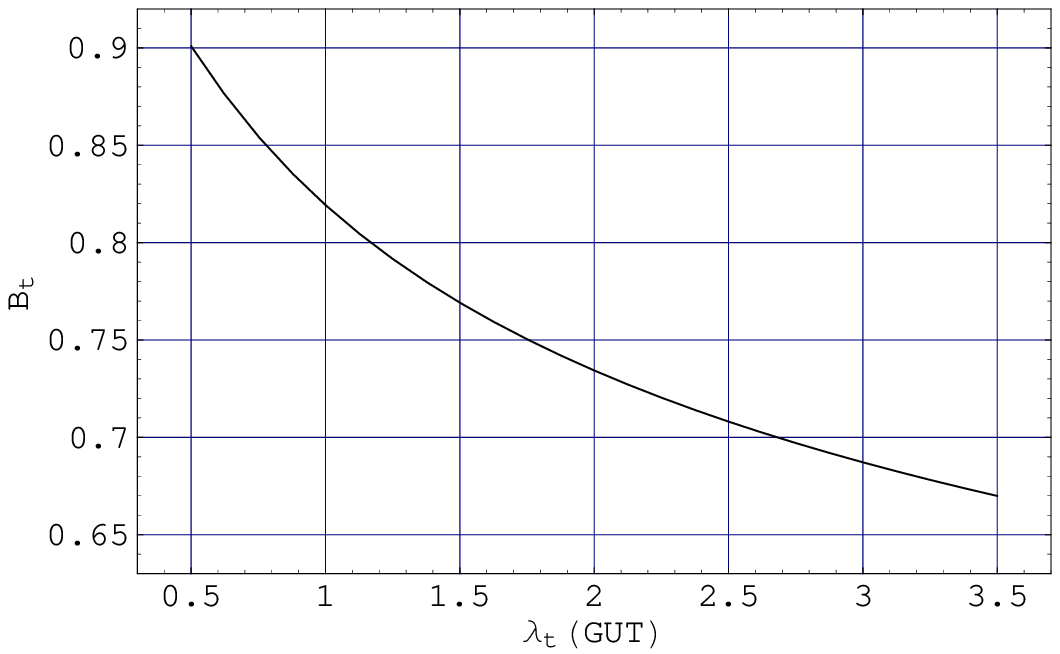,width=19em}%
\vspace*{-2ex}
\caption{$B_t$ as a function of $\l_t$.\label{fig:bt}}%
}

Finally, the factor $B_t$ is a function of $\l_t$, encoding the
nonlinearity in the ``top'' renormalization due to its near
Pendleton-Ross supersymmetric fixed point. We obtain
$B_t|_{\l_t\!=\!0.5}=0.901$, $B_t|_{\l_t\!=\!3.5}=0.670$.  Near lower
$\l_t$ the dependence on $\l_t$ is stiffer, and in power 6 it allows a
quite precise determination of $\l_t$ from the experimental value of~$m_t$.
Indeed, one can estimate the factor $B_t$ after determining $\l_t$
from $m_t$, once one notices that $m_t$ is almost independent of $\tb$
in its moderate range, i.e.\ $\sin{\b}\simeq0.99$ within 0.5\%, and
that $\l_t$ turns out to lie near 0.5, so that $B_t$ is sufficiently
dependent of $\l_t$.  Taking into account the uncertainty in~$m_t$ one
finds $\l_t = 0.536\mbox{--} 0.809$ and $B_t=0.8734 \pm 2.5\%_{m_t}$.

At this point, identities~(\ref{eq:yukunif}) can be rewritten in the
following form involving only low energy fermion masses:
\bea
\label{eq:dspred}
m_d m_s&=&\frac{m_e m_\mu m_\tau}{m_b}\, \frac{R_d^3
  \eta_b\eta_{uds}^2}{R_e^3 \eta_{e\mu}^2\eta_\tau} B_t\\[1ex]
\label{eq:upred}
m_u &=&\frac{m_e m_\mu m_\tau}{m_c m_t}\, \frac{R_u^3
  \eta_c\eta_{uds}}{R_e^3 \eta_{e\mu}^2\eta_\tau} (\tb)^3 B_t^{12}
\eea
and this implies the following predictions for $m_dm_s$ and $m_u$:
\bea
m_dm_s&=&880 \,\MeV^2\pm 2\%_{m_t}\pm4\%_{m_b}\pm25\%_{\a_3}\label{eq:ds}\\[1ex]
m_u&=&\bigg(\frac\tb7\bigg)^3 1.95\,\MeV\pm20\%_{m_t}\pm10\%_{m_c}\pm15\%_{\a_3}\,.\label{eq:mu}
\eea
Using a mean value of 19.5 for the known ratio $m_s/m_d$, the first equation
gives the central values
\be
 m_d=6.7\,\MeV\,,\qquad m_s=131\,\MeV\,,
\ee
that fit well the data, with $m_s$ a bit high, within 1$\sigma$ of the
experimental range. This means that experimentally the D, E
determinants are unified at GUT scale within 1-$\sigma$.\footnote{The
same analysis carried out in the non-supersymmetric Standard Model
gives an even better agreement of the D, E determinants (within 2\%)
at its 1-2 unification scale $M_{\mbox{\scriptsize GUT-12}}\simeq10^{13}\,\GeV$.
However, even ignoring the lack of simple unification of the three
gauge coupling constants, the requirement of the U determinant to
match with D, E requires an extension to models of the 2HDM type, with
the relative $\tb$ parameter again in the moderate range.}

The prediction for $m_u$ carries a dependence on \tb, that can be
used to derive a prediction for it:
\be
\tb=7.3\,\pm1.3_{m_u}\,\pm0.5_{m_t}\,\pm0.2_{m_c}\,,
\ee
where we used the 1$\sigma$ range for $m_u=(3\pm
1.5)\,\MeV$~\cite{pdg}. This prediction lies exactly in the favored
range determined from the recent analysis in various MSSM
scenarios~\cite{pdg,higgs}.

An other prediction for $\tb$ may be derived by combining the two
above and assuming for $m_u/m_d$ and $m_s$ their experimental
ranges. One gets:
\bea
\tb&=&\bigg(\frac{m_u}{m_d}\,\frac{m_cm_t}{m_sm_b}\bigg)^{1/3}
      \frac{R_u}{R_d}\bigg(\frac{\eta_b\eta_{uds}}{\eta_c}\bigg)^{1/3}
      B_t^{-11/3}\0\\[1ex]
&=&10\,\pm1.3_{m_u/m_d}\,\pm1_{m_s}\,\pm 1_{m_t}\,\pm0.5_{\a_3}\,.
\eea
Of course this prediction tells what is the value of \tb\ once the
model has been solved and $m_u/m_d$, $m_s$ and $m_t$ are
determined. However we have found that the dependence on $\a_3|_Z$
cancels almost perfectly in both the ratios involving the $R$'s and
the $\eta$'s, and enters only through the $B_t$ dependence
($<\!4\%$). This is a remarkable fact, considering that for individual
masses the uncertainty in $\a_s|_Z$ generates the dominant errors in
their predictions.

In addition, and equally remarkably, the even larger dependence on the
supersymmetry breaking scale also cancels almost exactly in the ratio
$R_u/R_d$, and enters just through the $B_t$ factor, where it can be
estimated as a small decrease of 3\% when $M_{\rm SUSY}$ is raised from
$167\,\GeV$ to $1\,\TeV$.

\section{Model analysis}
\label{sec:anal}

One may take the decomposition in rank-one projectors described until
here as an ansatz, and verify that it can explain all fermion masses
and mixing angles. To this aim we will write the mass matrices in
terms of a number of independent parameters, and show how to determine
them from the known masses, mixing angles, and CP violation phase.

First of all let us parametrize the symmetric flavour projectors $\hat
P_n$ in terms of the three triplets $p_n$, and choose a flavour basis
to write the projectors as follows:
\bea
\label{eq:dir}
  \hat\l_f = \l\Big[\e_f\hat P_1\,+\hat P_2\,+\e_f^{-1}\hat P_3\Big]&=&\l\left[
 \e_f  \matr{1&a&b\\a&a^2&ab\\b&ab&b^2}
+      \matr{0&0&0\\0&1&c\\0&c&c^2}
+\e_f^{-1} \matr{0&0&0\\0&0&0\\0&0&d^2}
                           \right]\0\\
&=&\l
\arraycolsep=.8em
\matr{\e_f& a\,\e_f&b \,\e_f \\[1ex]
a\,\e_f & 1+a^2\e_f& c+ab\,\e_f\\[1ex]
b\,\e_f & c+ab\,\e_f & d^2/\e_f+c^2+b^2\e_f}\,,
\eea
where we have rotated globally all the $p_n$ vectors in flavour space
and rescaled $\e_f$, $\l$ to set $p_1=(1,a,b)$, $p_2=(0,1,c)$,
$p_3=(0,0,d)$. Eliminating also the irrelevant phases, we end up with
the following 15 parameters domain: $\l,a,d\in {\mathbb R}^+$, $
b,c,\e_u,\e_d,\e_e,\e_{\nu_D}\in{\mathbb C}$. Because of the SO(10)
relations~(\ref{eq:soten}) two $\e$ may be eliminated and we remain
with \emph{11 real parameters}. This form should account for all the
Dirac couplings of charged fermions and neutrinos.

For the Majorana mass matrix of neutrinos, in view of~(\ref{eq:Lnu}),
we have a similar form, but involving $\e_{\nu_L}$ in place of $\e_f$
and with $\a$ appearing in the $d^2/\e_{\nu_L}$ term in the 3-3
entry. The neutrino sector thus adds \emph{4 real parameters} to the model.

\subsection{Leading order}
\label{sec:firstorder}

The yukawa sector can be analyzed quite completely by computing the
leading order expressions for eigenvalues and mixing angles. For
charged fermions:
 \bea
\label{eq:yukup}
 \l_u\simeq\l\,|\e_u|\,,\ \ \qquad\quad&\quad&\l_c\simeq\l\,,\quad\ \ \ \ \ \
 \qquad\qquad\l_t\simeq\l\left|\frac {d^2}{\e_u}\right|
\\
 \l_d\simeq\l\frac{|\e_d|}{|1+{ a^2\e_d}|}\,,\quad&\quad&\l_s\simeq\l|1+{
 a^2\e_d}|\,,\quad\qquad\l_b\simeq\l\left|\frac {d^2}{\e_d}\right|
\\
 \l_e\simeq\l\frac{|\e_e|}{|1+{
 a^2\e_e}|}\,,\quad&\quad&\l_\mu\simeq\l|1+{a^2\e_e}|\,,\quad\qquad\l_\tau\simeq\l\left|\frac {d^2}{\e_e}\right|
 \eea
 \vspace*{-1em}
 \be
 \t_{12}\simeq-\frac{a\e_d}{1+a^2\e_d}\,,\qquad\t_{23}\simeq 
 \frac{\eta_d}{d^2}\e_d\,,\qquad \t_{13}\simeq \frac b{d^2} \e_d^2\,,\qquad
  \gamma_{\rm CP}\simeq\arg\left(\frac{b (1+a^2\e_d)}{\eta_d}\right),
 \ee
 where $\eta_d=(c+ab\, \e_d)$. Note also that since $\e_u\ll\e_{d,e}$,
 the quark mixing angles are well approximated by the ``down''
 sector. The analysis proceeds with the following steps:
\begin{itemize}
\item First we note from the third generation that, to respect
  $\l_b/\l_\tau\simeq1$, one should have $|\e_e|\simeq|\e_d|$ as
  already discussed.  This relation holds within 10\%, but is
  nevertheless important for the following first order calculation.

\item Then, if one notes that ignoring the corrections the second
  generation Yukawa are unified $\l_{c,s,\mu}\simeq\l$, on the other
  hand the RG invariant relation $s_{12}\simeq\sqrt{\Md/\Ms}$ tells us
  that $|a^2\e_d|\simeq1$, so that $a$ should actually be quite large,
  $|a|\sim|\e_d|^{-1/2}$ ($\simeq7$, see later) and the corrections
  can not be neglected.

\item On the contrary, one sees that the required split of s,$\mu$
  Yukawa is produced by the factors $A_{d,e}=|1+a^2\e_{d,e}|$ once the
  sign (the phases) of $a^2\e_e$ and $a^2\e_d$ are nearly opposite,
  $a^2\e_e\simeq-a^2\e_d\simeq\mbox{e}^{0.5i}$, so that
  $|A_e|\simeq2$, $|A_d|\simeq0.5$. For the first generation the split
  goes (correctly!) in the opposite direction.  Therefore, in this
  approach, the largeness of $a$, following from the Cabibbo angle,
  explains also the deviation from exactly family hierarchical masses
  of $e$, $d$, $\mu$, $s$. This fact was observed in~\cite{br}.

\item From other known mass ratios one can then determine
  $|\e_{e,d}|\simeq1/55$, $a\simeq7.5$, $|\e_u|\simeq1/600$,
  $d^2\simeq0.5$, $\tb\simeq7.3$.

\item Further insight can be gained by looking at the other quark
  mixing angles. From: \be
  \label{eq:angles}
  \theta_{13} \simeq |b\,\e_d^2/d^2|=3\cdot 10^{-3} \,,\qquad
  \theta_{23} \simeq |\eta_d\e_d/d^2|=4\cdot 10^{-2} \,, \ee where
  $\eta_d=(c+ab\,\e_d)$, one derives $|b|\simeq 5$,
  $|c|\simeq0.5\mbox{--}1.5$. Therefore the consequences of $\theta_{13}$
  are similar to the effect of the Cabibbo mixing: while there we got
  that numerically $a\simeq\e_d^{-1/2}$, now we see that also
  $b\simeq\e_d^{-1/2}$ is large, of the same order. This does not
  introduce significant correction to the third-generation eigenvalues
  but, below, will help neutrino to have a maximal atmospheric mixing.

\item A phase can be determined from the CKM CP-violating phase
  $\gamma_{\rm CP}$:\footnote{We assume that no other contribution to
  $\gamma_{\rm CP}$ is introduced between GUT and our low energy scale. In
  such a case this formula should be modified accordingly.}
  \be
  \gamma_{\rm CP}\simeq\arg\bigg(\frac{b(1+a^2\e_d)}{\eta_d}\bigg)=1.02\,,
  \quad\Longrightarrow\quad
  \arg(b)-\arg(\eta_d)=1\pm1=\Big\{\ba{ll}0\\2\ea\,,
  \ee
  where we had from the first two steps that $\arg(1+a^2\e_d)\simeq\pm
  1$, the sign depending on complex conjugation of $\e_d$.  Fixing in
  this way the relative phase of $b$ with $\eta_d$ also fixes the
  modulus of $c=\eta_d-ab\,\e_d$.
  The two branches correspond to $|c|\simeq0.2$, $|c|\simeq1.5$. In the
  first branch, $c$ and $b$ are approximately orthogonal, in the
  second they are approximately opposite.

\item Two remaining free parameters can be taken to be
  $|\e_e/\e_d|$, $\arg c$, and one can be eliminated (using second
  order corrections to eigenvalues) from the precise ratio
  $\l_b/\l_\tau \simeq [|\e_e/\e_d|+2\Re(\e_dc^2/d^2)]$.

\end{itemize}
Summing up, the leading order analysis for charged fermions shows a
one parameter family of solutions, with two branches.  The flat
direction can be taken as a combination of $|\e_e/\e_d|$
with the common phase of $b$, $c$.

The precise leading-order solution for all parameters is not
conclusive since next-to-leading corrections and contributions to CKM
from the ``up'' sector are 10\% in magnitude, larger than experimental
errors; hence a numeric fit is necessary and will be described below.
Nevertheless the model proves capable of accounting for all the
charged fermion masses and mixings, giving a mechanism for linking the
deviations $A_{e,d}$ with the Cabibbo angle $\t_{12}$.

We can now comment on the solution found from the charged fermions so
far. Collecting the $p_n$ vectors:
\bea
p_1=(1,\,7,\,5)\,,\qquad &p_2\simeq(0,\,1,\,0.2\mbox{--}1.5)\,,& \qquad p_3\simeq(0,\,0,\,0.7)\,,
\label{eq:vectors}
\eea
we observe that the they tend to lie in the 2-3 plane.  Exact
planarity is broken to order $1/8\simeq\sqrt{\e_d}$, as can be seen
from the first 1, with respect to the modulus of its vector
$|p_1|\simeq 8.5$.  The reason for this is to be tracked in the
magnitude of the quark mixing angles that require $a$ and $b$ to be
large while $c$ and $d$ are of order one.\footnote{Aa a side note, the
value found for $d^2\simeq0.5$ tells that the eigenvalues in the
``up'' sector as from line~(\ref{eq:yukup}) do not strictly follow an
exact hierarchy as in~(\ref{eq:up}). A factor of 0.5 will decrease,
with respect to precise hierarchy, either the top mass, which would be
unacceptable, or the up quark mass, which is well tolerable. We will
see also from the numerical solution that low $m_u$ is a prediction of
this ansatz (see e.g.\ figure~\ref{fig:result}).}  In terms of the
reciprocal vectors $q_{1,2,3}$ that appear in the model II realization
in section~\ref{sec:realizationII}, quasi-planarity shows up as
quasi-alignment. We have
\be
q_1=(1,0,0)\,,\qquad q_2=(7,1,0)\,,\qquad q_3\simeq(7,1,1)\,.
\ee

We find this pattern of quasi-planar or quasi-aligned triplets a nice
hint for the realization of in the flavour sector of the theory, where
three flavour triplets scalar fields acquire a VEV that generates the
$p_n$ vectors. We leave this possibility for future model building;
see e.g.~\cite{disoriented} for realizations of disoriented VEVs from
a potential.

We observe also that if one may be concerned with the moduli of $p_n$
not being equal and of order 1, (they are in fact $|p_1|\simeq 8$,
$|p_2|\simeq1.5$, $|p_3|\simeq0.7$), they can be brought to be in the
range 1--2 by rescaling all the $\e_f$ by a factor of 10. With this
normalization the hierarchy of eigenvalues is due partially to the
$\e_f$ and partially to the $p_n$ vectors having hierarchic entries,
while the normalization that we adopted above corresponds to an
eigenvalue hierarchy that comes only from the $\e_f$'s.  The choice
may be varied depending on the theoretical realization in the flavour
sector, that should be the subject of a further study and goes beyond
the scope of this work.

\subsection{Neutrino analytical}

From the light neutrino mass matrix~(\ref{eq:Lnu}) and using the
explicit parametrization given in~(\ref{eq:dir}), we can
rewrite:
\vspace*{-1ex}
$$
\l_{\nu_L}\propto
\matr{|\e_\nu/\bar A_\nu|&|a\e_\nu/A_\nu|&|b\e_\nu/A_\nu|\\
      |a\e_\nu/A_\nu|    &  1        &\rlap{$\rho$}\mbox{\phantom{$\rho^2 +\sigma$}}\\
      |b\e_\nu/A_\nu| & \rho      & \rho^2 +\sigma}.
$$
with $A_\nu=(1+a^2\e_\nu)$,~~$\rho= (c+ab\e_\nu)/A_\nu$,~~$\sigma=(d^2\a+(b-ac)^2\e_\nu^2)/A_\nu$.

In this matrix, we only have two free complex parameters $\e_\nu$ and
$\a$, and the common phase of $b$, $c$ that was a flat direction of
the charged fermions.  However it is easy to see by homogeneity that
this latter phase is irrelevant since it can be eliminated using the
phase of $\a$.  Therefore, we have just two complex parameters
$\e_\nu$ and $\a$, and also the neutrino sector is quite constrained.

Because of the large neutrino mixings and of the small but important
uncertainties in the parameters determined from the charged fermions,
an exact analytical decomposition of this matrix is quite difficult.
Nevertheless some useful observations can be made:
\begin{itemize}
\item First, since the ratio of the 1-1 with the 1-2, 1-3 entries
is small $\simeq1/7$, from the known pattern of mixing angles it
follows that this matrix predicts  direct and non degenerate neutrinos:
$m_{\nu1}/m_{\nu2}\simeq1/4$ so that $m_{\nu2}/m_{\nu3}\simeq1/5$.

\item Then, because the 1-2 entry is larger than the 1-3 one
(1-2/1-3 $\simeq1.5$--2), then $\t_{e3}$ cannot be zero.

\item An estimate of the angles from the 2,3 sector can be given
first in the form of a correlation between them, using the parameters
found from the charged sector and after the requirement of maximal
$\t_{\rm atm}$:
\be
\frac{\t_{e3}}{\t_{\rm sol}} = \frac{m_{\nu2}}{m_{\nu3}}\,
\Big|\frac{1+R\,{\rm e}^{i\delta}}{1-R\,{\rm e}^{i\delta}}\Big|\,,
\qquad
\mbox{with }\ R=\frac{b}{a}\simeq\frac{\t_{13}^{\rm quark}\l_b}{\t_{12}^{\rm quark}\l_s}
\simeq \frac1{1.6 \thicksim 2.1}\,,
\ee
where $\delta=\arg \rho$. After imposing the correct neutrino
hierarchy to solve for $\e_\nu$, we can estimate more explicitly:
\be
\t_{\rm sol}\simeq 30^\circ \Big|1-R\,{\rm e}^{i\delta}\Big|\,,\qquad 
\t_{e3}\simeq 6^\circ \Big|1+R\,{\rm e}^{i\delta}\Big|\,.
\ee
Therefore, choosing ${\rm e}^{i\delta}$ purely imaginary or just
slightly negative, we see that the right $\t_{\rm sol}\simeq32^\circ$ is
accommodated, with a prediction for $\t_{e3}\simeq5^\circ$.

The precise numbers are sensible to the precise values of the quark
mixing angles and CP-phase, as well as to the neutrino hierarchy. In
practice $\t_{e3}$ can be brought reach the extrema of the range
$\t_{e3}\simeq1^\circ\mbox{--}7.5^\circ$, by stretching within
1$\sigma$ the angles and the neutrino hierarchy. This means that the
model predicts a nonzero $\t_{e3}$ inside the present 2$\sigma$ range
(=0$^\circ$--9.6$^\circ$). With more precise experimental values of
CKM angles and neutrino hierarchy, also the prediction for $\t_{e3}$
will become narrower.

\item Finally, it can be seen that among the four parameters a flat
direction emerges, along which the angles and hierarchy do not
vary. It directly corresponds to the leptonic CP violation phase, that
is thus left unpredicted by this ansatz, until angles and hierarchy
will be known to better accuracy. Also one Majorana phase varies along
this flat direction.
\end{itemize}

Summing up, in the neutrino sector the model predicts hierarchic
neutrinos with nonzero $\t_{e3}$ in the present range, and there is a
useful conspiracy from the charged fermions sector to allow the right
neutrino pattern.  A flat direction emerges as a combination of the
four real parameters, and has to be added to the one found in the
charged leptons sector.  Therefore the whole model effectively takes
advantage of just 13 of the 15 real parameters to reproduce the known
data.

For a precise test we performed a numerical analysis, where we fitted
neutrinos together with charged fermions.  The $\t_{e3}$ mixing, that
we have shown to lie automatically inside its allowed range, will not
be used as a fit parameter but rather displayed as a model prediction.

\subsection{Numeric fit and results}

The fit was performed using the 15 parameters (11 charged fermions, 4
neutrino) against 20 tests for all known data~\cite{pdg,vs}.  The best
fit results are shown in tables~\ref{tab:fit1} and~\ref{tab:fit2}, for
models I and II. Each model has solutions in two branches, and model I
shows difficulties in fitting the ``up'' sector ($m_t$ is too low, and
$m_c$ too high). Model II instead can fit almost perfectly all data,
albeit predicting a bit low $m_u$ and a bit large $m_s$ as implied by
the predictions described in section~\ref{sec:detunif}.

The dependence on the supersymmetry breaking scale is the major source
of theoretical uncertainty, and we have estimated its effect by
performing the fit also with $M_{\rm SUSY}=1\,\TeV$ and including mixed RG
evolution.  Raising the SUSY scale leads to better results in both
models, but more appreciably in model II where all tension with
experimental data disappears, and unification of determinants works
perfectly.

\begin{figure}[p]
\centerline{\epsfig{file=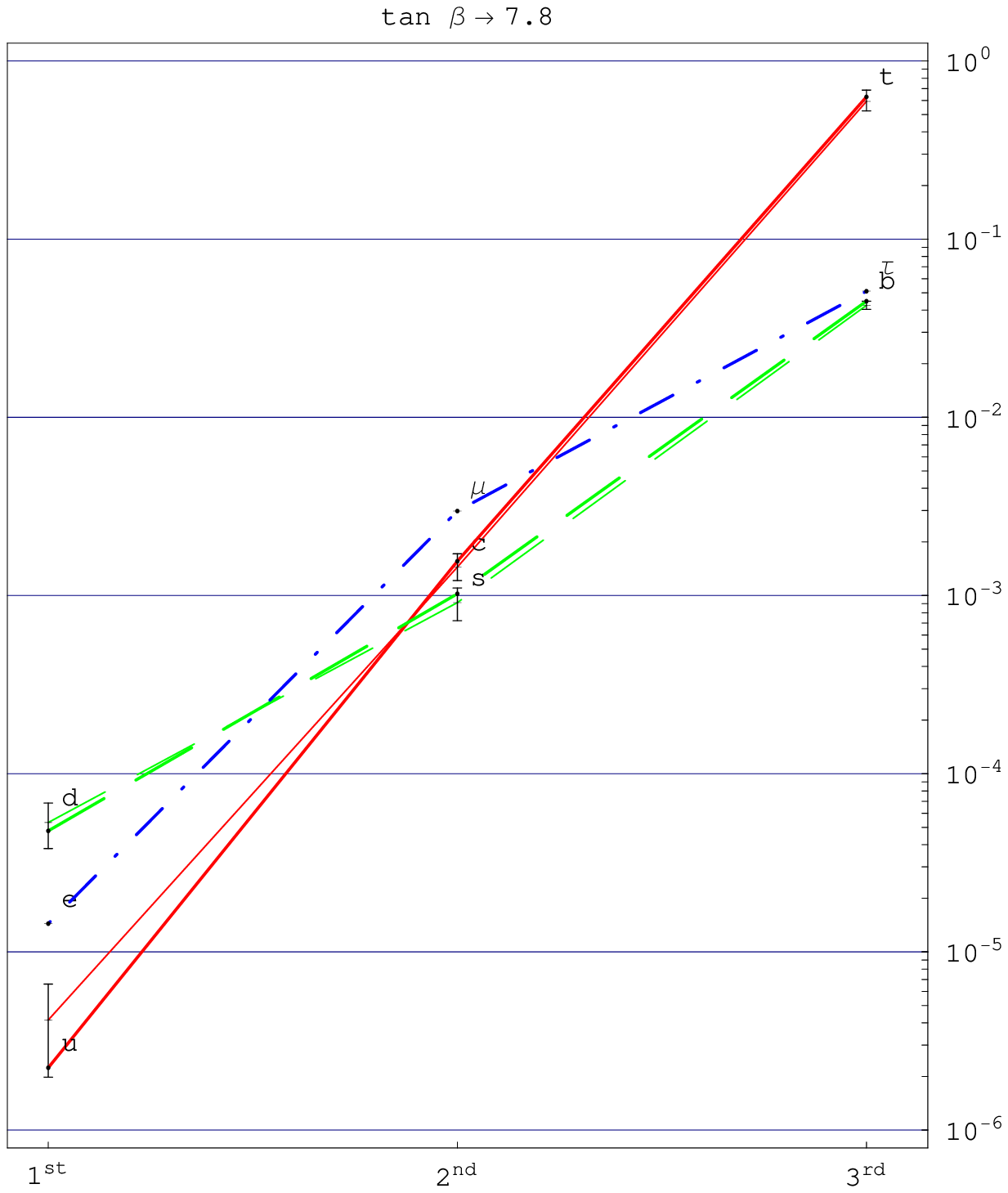,height=46em}}
\caption{Illustration of yukawa couplings at GUT scale for the three
  families of charged fermions, resulting from the best fit in model
  II for $M_{\rm SUSY}=167\,\GeV$, compared to the values calculated from
  experimental masses (as shown in figure~\protect\ref{fig:abc}). Thin lines
  connect yukawa calculated from experimental masses, thick lines
  correspond to the best fit result.}
\label{fig:result}
\end{figure}

For model II we also illustrate the best fit Yukawa couplings in
figure~\ref{fig:result}.

A number of observations from the fit results in model II can be made:
\begin{itemize}
\item Apart from $m_u$ and $m_s$ with lower $M_{\rm SUSY}$ that are
         predicted in the $1\sigma$ range, all data are accomodated
         without any tension.  Within the same total deviation one may
         choose even to adjust better either $m_s$, $m_u/m_d$ or $m_c$
         and $m_t$.
\item A flat direction and the two branches $|c|\simeq0.2,1.5$
         correctly emerge from the fit. In the second branch all
         complex phases of all parameters are almost aligned, hinting
         for a model with aligned $\e_f$'s and real parameters, with a
         reduction of the total number of real parameters to 10.
\item In the lepton sector, the mixing from terrestrial neutrinos
         is nonzero: $\t_{e3}\simeq2^\circ,6^\circ$, respectively
         in the two branches.  
\item Neutrino come out hierarchic and non degenerate
         $m_{\nu}\simeq(0.0045,\, 0.01,\, 0.048)\,\eV$, and
         right-handed neutrino masses are of the order ($10^7$,
         $10^9$, $10^{11}$) \GeV.
\item The leptogenesys asymmetry appears to have the correct sign
         $\e_1\simeq-3\cdot10^{-9}$ at least in some branches, though
         (due to small RH neutrino masses) the value itself is too
         small for the usual thermal scenario, and it requires a
         specific assumption on the non-thermal production of the
         heavy neutrino initial abundances.
\end{itemize}

\begin{table}[p]
\vspace*{.8ex}
\small
\centerline{%
$
\def\arraystretch{1.2}
\arraycolsep=.8em
\begin{array}[t]{|c|c||c|c||c|c|}
\hline
\multicolumn{6}{|c|}{\mbox{MODEL I}}\\
\hline
             \multicolumn{2}{|c||}{\,}
            &\multicolumn{2}{c||}{M_{\rm SUSY}=m_t}
            &\multicolumn{2}{c|}{M_{\rm SUSY}=1\TeV}\\ 
& \mbox{exp}
            &\mbox{$1^{\rm st}$ branch}&\mbox{$2^{\rm nd}$ branch}
            &\mbox{$1^{\rm st}$ branch}&\mbox{$2^{\rm nd}$ branch}\\
\hline
\multicolumn{6}{c}{Fit}\\
\hline
m_t               & 167\pm5           & {\bf 147}      & {\bf 148}      &  {\bf 155}   & {\bf 155}     \\
m_b               & 4.25\pm0.15       &   4.36         &   4.38         &   4.37       &   4.37        \\
m_c               & 1.25\pm0.10       & \ {\bf 1.49 }\ & \ {\bf 1.47 }\ & {\bf 1.46}   & {\bf 1.45}    \\
m_s               & 105\pm25          & {\bf 132}      & 119            & 128          & 116           \\
m_s/m_d           & 19.5\pm2.5        &  21.1          &  21            &  21.8        & {\bf 22.2}    \\
\mbox{Ellipse}    & 23\pm2            &  22.1          &  22.6          &  22.5        &   22.9        \\
m_u/m_d           & 0.5\pm0.2         & {\bf 0.26 }    & {\bf 0.27 }    & {\bf 0.26 }  & {\bf 0.25}    \\
m_u\!\!+\!m_d     & 8.5\pm2.5         &   7.6          &   8.0          &   7.3        &    6.5        \\
\hline				      					 	    		      
\theta_{12}       & 12.7\pm0.1        & 12.75          &  12.75         & {\bf 12.81}   &  {\bf 12.83}  \\
10^2\,\theta_{23} & 4.13\pm0.15       & 4.11           &  4.12          & 4.11         &  4.12         \\
10^3\,\theta_{13} & 3.67\pm0.47       & 3.82           &  3.79          & 3.82         &  3.79         \\
\sin\!2\b_{\rm CP}    & 0.736\pm0.05      & 0.735          &    0.734       & 0.730        &    0.734      \\
\gamma_{\rm CP}       & 1.03\pm0.22       & 0.93           &    1.05        & 0.93         &    1.02       \\
\hline				      					 	    	     	      
\theta_{\rm sol}      & 32\pm3            & 31.            &  32.           & 31.0         &  32.          \\
\theta_{\rm atm}      & 45\pm7            & 45.3           &  45.           & 45.3         &  45.          \\
h_\nu             & 0.18\pm0.07       & 0.165          &  0.180         & 0.171        &  0.185        \\
\hline		  		     					 	    		      
\hline				     					 	    		      
\chi^2_{\rm dof}      &                   &    0.68        &  0.53          & 0.49         &  0.43         \\
\hline									 
\multicolumn{6}{c}{Predictions}\\					 		      
\hline									 		      
m_d\,(\MeV)           & 6\pm2         &      6.0       &   5.4          &   5.9        &   5.25        \\
\theta_{e3}\,(^\circ) &0\mbox{ -- }9.6&    1.8         &  6.6           & 1.8          &  7.2          \\
m_{\nu1}\,(\eV)       &               &   0.0029       &  0.0026        &  0.0029      &  0.0028       \\
m_{\nu3}\,(\eV)       &               &   0.052        &  0.048         &  0.052       &  0.048        \\
m_{ee}\,(\eV)         &               &   0.0007       &  0.0009        &  0.0007      &  0.0009       \\
\e_1                  &               &   +2\,10^{-12} &  +6\,10^{-12}  & +2\,10^{-12} & +2.7\,10^{-12}\\
\hline				          				 			      
\tb                   &               &    1.21        &  1.23          & 1.13         & 1.14          \\
\hline												     
\end{array}
$
}
\vspace*{1.5ex}
\caption{Result of best fit against the known experimental ranges, for
  the two branches of model I, varying also the supersymmetry breaking
  scale between the ``top'' scale $167\,\GeV$ and $1\,\TeV$.  In bold
  are shown the quantities that show a $1\sigma$ tension with
  experiment. Lepton masses are reproduced exactly and are not
  displayed.  We report also some predicted quantities, not used for
  fitting, but determined in the best fit solution. This model shows a
  evident difficulty in accommodating the ``up'' quark sector.
  \vspace*{-1ex}}
\label{tab:fit1}
\end{table}

\begin{table}[p]
\vspace*{.8ex}
\small
\centerline{%
$
\arraycolsep=.8em
\begin{array}[t]{|c|c||c|c||c|c|}
\hline
\multicolumn{6}{|c|}{\mbox{MODEL II}}\\
\hline
             \multicolumn{2}{|c||}{\,}
            &\multicolumn{2}{c||}{M_{\rm SUSY}=m_t}
            &\multicolumn{2}{c|}{M_{\rm SUSY}=1\TeV}\\ 
& \mbox{exp}
            &\mbox{$1^{\rm st}$ branch}&\mbox{$2^{\rm nd}$ branch}
            &\mbox{$1^{\rm st}$ branch}&\mbox{$2^{\rm nd}$ branch}\\
\hline
\multicolumn{6}{c}{Fit}\\
\hline
m_t               & 167\pm5           &   167.6           &   169.0           &   167.4      &   167.8       \\
m_b               & 4.25\pm0.15       &   4.40            &   4.45            &   4.32       &   4.38        \\
m_c               & 1.25\pm0.10       &   1.32            &   1.31            &   1.29       &   1.29        \\
m_s               & 105\pm25          & {\bf 134}         & {\bf 134}         & 127          & 126           \\
m_s/m_d           & 19.5\pm2.5        &  21.1             &   21.4            &  21.01       &   21.0        \\
\mbox{Ellipse}    & 23\pm2            &  21.5             &   22.1            &  21.46       &   22.2        \\
m_u/m_d           & 0.5\pm0.2         & {\bf 0.20 }       & {\bf 0.25}        & {\bf 0.21 }  &    0.32       \\
m_u\!\!+\!m_d     & 8.5\pm2.5         &   7.5             &   8.0             &   7.3        &    7.9        \\
\hline				      	            		       		    		      
\theta_{12}       & 12.7\pm0.1        & 12.74             &  12.77            & 12.72        &  12.75        \\
10^2\,\theta_{23} & 4.13\pm0.15       & 4.12              &  4.14             & 4.11         &  4.13         \\
10^3\,\theta_{13} & 3.67\pm0.47       & 3.77              &  3.77             & 3.77         &  3.76         \\
\sin\!2\b_{\rm CP}    & 0.736\pm0.05      & 0.730             &    0.734          & 0.73         &    0.734      \\
\gamma_{\rm CP}       & 1.03\pm0.22       & 0.97              &    1.07           & 0.98         &    1.06       \\
\hline				      	            		       		    	     	      
\theta_{\rm sol}      & 32\pm3            & 31.5              &  32.              & 31.4         &  32.          \\
\theta_{\rm atm}      & 45\pm7            & 45.1              &  45.              & 45.2         &  45.          \\
h_\nu             & 0.18\pm0.07       & 0.178             &  0.185            & 0.176        &  0.185        \\
\hline		  		      	            		       		    		      
\hline				      	            		       		    		      
\chi^2_{\rm dof}      &                   & 0.30              &  0.27             & 0.22         &  0.14         \\
\hline				      			     
\multicolumn{6}{c}{Predictions}\\     			     				      
\hline				      			     				      
m_d\,(\MeV)           & 6\pm2         &   6.4             &   6.3             &   6.1        &   6.0         \\
\theta_{e3}\,(^\circ) &0\mbox{ -- }9.6& 2.1_{-0.4}^{+2.2} & 6.4_{-5.5}^{+0.8} & 1.9          &  5.9          \\
m_{\nu1}\,(\eV)       &               &  0.0031           &  0.0030           &  0.0030      &  0.0044       \\
m_{\nu3}\,(\eV)       &               &  0.050            &  0.048            &  0.050       &  0.0481       \\
m_{ee}\,(\eV)         &               &  0.00068          &  0.00091          &  0.0007      &  0.00085      \\
\e_1                  &               & +3\,10^{-10}      &  -2\,10^{-9}      & +2\,10^{-10} & -3.3\,10^{-9} \\
\hline				      				     				      
\tb                   &               & 7.1_{-0.3}^{+0.5} & 7.8_{-0.5}^{+0.5} & 7.7          & 8.9           \\
\hline												     
\end{array}
$
}
\vspace*{1.5ex}
\caption{Result of best fit against the known experimental ranges, for
  the two branches of model II, varying the supersymmetry breaking
  scale between the ``top'' scale $167\,\GeV$ and $1\,\TeV$.  In bold
  are shown the quantities that show a $1\sigma$ tension.  For
  $\t_{e3}$ and $\tb$ we also indicate the ranges obtained, by
  stretching these quantities up to an additional 1$\sigma$ variation.
  It can be noted that raising the supersymmetry scale to $1\,\TeV$
  eliminates almost all tension, especially in the 2$^{\rm nd}$ branch.
  \vspace*{-1ex}}
\label{tab:fit2}
\end{table}

\clearpage

\section{Conclusions}
\label{sec:outlook}

We presented a predictive model of SU(3) flavour symmetry in the
context of SO(10) supersymmetric GUT that can account for all the
known fermion masses and mixings at less than 1$\sigma$ level,
including neutrino data. The model has been analyzed with a
leading-order analytical approach and with a complete numerical fit.

In the quark sector, it predicts 1$\sigma$ low $m_u$ and, without
raising the SUSY breaking scale, 1$\sigma$ large $m_s$; all the other
quantities are well accounted, including the CP phase.
In the neutrino sector the model predicts direct and hierarchical
neutrinos with non zero $\t_{e3}\simeq2^\circ,6^\circ$ correlated to
other quantities, but always within the present 99\% c.l.. Also the
correct sign of the leptogenesys asymmetry is an outcome of the
numerical analysis, even though its magnitude is one or two orders of
magnitude too low for the usual thermal leptogenesys scenario.

The SO(10) realization of this framework uses an ``universal'' seesaw
mixing with superheavy vector-like fermions to transfer their mass
matrices to the light fermions.  Together with adjoint Higgs fields,
it has been shown to suppress the dominant part of D=5 proton decay
and supersymmetric flavour changing effects, while solving the
doublet-triplet problem via the Missing VEV mechanism and allowing the
correct mass generation.

The flavour structure of the model is based on hierarchical
combinations of three rank-one projectors in flavour space, built with
three flavour triplets or sextets, that break the SU(3) symmetry. As
opposed to traditional unification schemes, this model does not unify
some Yukawa matrices at GUT, but realizes in an interesting way the
exact unification of their determinants.

A quite robust prediction for $\tb\simeq8$ in the moderate, favored,
range is also consequence of this kind of unification.

The solution emerging from the analysis points to an interesting
configuration of quasi-aligned flavour triplets, that suggests a
precise pattern of SU(3) breaking. We leave this analysis for further
work.  Finally, the intertwining between flavour and SO(10) Higgs
fields also suggests that a similar mechanism in the context of higher
unification groups may lead to a framework as successful as this one.

\acknowledgments 

We would like to acknowledge interesting discussions with A. Rossi,
G. Senjanovic and F. Vissani.  This work was partially supported by
the MIUR grant under the Projects of National Interest PRIN 2004
``Astroparticle Physics''.

\end{document}